\documentclass[conference]{IEEEtran}

\usepackage{fancyhdr}
\usepackage[square,numbers,sort&compress]{natbib}
\usepackage{tabularx, color, colortbl}
\usepackage{graphicx, subfigure}
\usepackage{amsmath, amssymb}
\usepackage{algorithm}
\usepackage{algpseudocode}
\usepackage{listings}
\usepackage[noblocks]{authblk}
\usepackage{mathtools}
\usepackage{todonotes}
\usepackage{framed}
\usepackage{subfig}
\usepackage{soul}
\usepackage{verbatim}
\usepackage{url}
\usepackage{booktabs}
\usepackage{flushend}
\usepackage{comment}
\usepackage{multirow}

\usepackage{cleveref}
\crefname{section}{§}{§§}
\Crefname{section}{§}{§§}

\newcommand{\rbt}[1]{{#1}} 

\begin{document}

	\title{RLScheduler: An Automated HPC Batch Job Scheduler Using Reinforcement Learning}
	\author[1]{Di Zhang} \author[1]{Dong Dai}
	\author[2]{Youbiao He} \author[2]{Forrest Sheng Bao}
	\author[3]{Bing Xie}
	\affil[1]{Computer Science Department, University of North Carolina at Charlotte,
		\{dzhang16, ddai\}@uncc.edu}
	
	\affil[2]{Computer Science Department, Iowa State University, \{yh54, fsb\}@iastate.edu}
	
	\affil[3]{Oak Ridge Leadership Computing Facility, Oak Ridge National Laboratory. xieb@ornl.gov}
	
	\maketitle

	\begin{abstract}
	    Today's high-performance computing (HPC) platforms are still dominated by batch jobs. Accordingly, effective batch job scheduling is crucial to obtain high system efficiency.
	    Existing HPC batch job schedulers typically
	    leverage heuristic priority functions to prioritize and schedule jobs. 
	    But, once configured and deployed
	    by the experts, such priority functions can hardly adapt to the changes of job loads, optimization goals, or system settings, potentially leading to degraded system efficiency \rbt{when changes occur}.
	    To address this fundamental issue, we present RLScheduler, an automated HPC batch job scheduler built on reinforcement learning. RLScheduler relies on minimal manual interventions or expert knowledge, but can learn high-quality scheduling policies via its own continuous `trial and error'. 
	    We introduce a new kernel-based neural network structure and trajectory filtering mechanism in RLScheduler to improve and stabilize the learning process.
	    Through extensive evaluations, we confirm that
	    RLScheduler can learn high-quality scheduling policies towards various workloads and various optimization goals with relatively low computation cost. Moreover, we show that the learned models perform stably even when applied to unseen workloads, making them practical for production use.
	\end{abstract}

	\section{Introduction}
	\label{sec:intro}
	\begin{table*}[t!]
    \centering
    \caption{Description of job attributes.}
    \label{tab:attribute}
    \begin{tabular}{c|c|l}
    \hline
    \hline
    \textbf{Name} & \textbf{Symbol} & \textbf{Description}\\
    \midrule
    \textit{Job ID}  & $id_t$ &  the id of job\\
    \textit{User ID} & $u_t$ & the user's ID\\
    \textit{Group ID} & $g_t$ & the group's ID\\
    \textit{Executable Id} & $app_t$ & ID of the job's executable file\\
    \textit{Submit Time}     & $s_t$ & job submission time \\
    \textit{Requested Processors}    & $n_t$ &  the number of processors that a job requests.\\
    \textit{Requested Time} & $r_t$ & job's runtime estimation (or upper bound) from users\\
    \textit{Requested Memory} & $m_t$ & the requested memory per processor\\
    \hline
    \hline
    \end{tabular}
\end{table*}

Today's high-performance computing (HPC) platforms are still dominated by batch jobs. On \rbt{such} a platform, jobs are submitted to a centralized job scheduler via job scripts and wait in a job queue until the scheduler allocates the requested resources for them to execute. Once start, the jobs will run til finish, or fail, or get killed, in a batch way~\cite{Feitelson1997TheoryScheduling}.

A batch job scheduler is designed to schedule jobs to obtain an optimization goal (or called metrics), such as {\it maximizing resource utilization}, {\it maximizing job throughput}, or {\it minimizing job wait time}, etc. Theoretically, batch job scheduling is NP-Hard~\cite{ullman1975np}. In practice, the HPC schedulers make scheduling decisions via heuristic \textit{priority functions}, which assign each job a priority based on its attributes.

In the context of batch job scheduling, the priority functions have been extensively studied~\cite{tang2009fault,Carastan-Santos2017ObtainingLearning,Xhafa2010ComputationalProblems,Mu2001UtilizationBackfilling,agarwal2006heuristics,akyol2007review,al2012power,yoo2003slurm,staples2006torque,nitzberg2004pbs}. In particular, some functions rely on a single job attribute, such as submission time (First Come First Server, FCFS) or job duration time (Shortest Job First, SJF)~\cite{pinedo2012scheduling}. Some compute priorities based on multiple job attributes~\cite{yoo2003slurm,staples2006torque,nitzberg2004pbs}. Recently, researchers proposed to use advanced algorithms, such as utility functions~\cite{tang2009fault} or machine learning techniques~\cite{Carastan-Santos2017ObtainingLearning}, to build priority functions. \rbt{A more detailed description about these schedulers and their priority functions can be found in Table~\ref{tab:funcs} in \S\ref{sec:eval}.}

However, no matter how a priority function is constructed (e.g., via careful workload analysis or yearly experts experiences), the forementioned schedulers share the same drawback: it is fixed and cannot automatically adapt to the variations in the target environment. On typical HPC platforms, job workloads may shift month by month and the optimization goals may also vary across time. For instance, when a cluster is deployed initially, system administrators may set the goal as high resource utilization and later change it to low average waiting time for addressing user interests. 

Manually tuning priority functions towards changing workloads or optimization goals is possible, but tedious and error-prone even for the most experienced system administrators.
Alternatively, an automated strategy would be more attractive. This motivates us to explore reinforcement learning (RL) methods~\cite{kaelbling1996reinforcement, Sutton2018ReinforcmentLearning} in batch job scheduling. Ideally, an RL-based job scheduler will adapt to the varying job load as RL can continuously learn from trial-and-error as the load varies; the scheduler will also adapt to various optimization goals as RL can 
{automatically} learn  the `best' policies for given rewards without manual intervention.

However, in practice, several key questions need to be answered before using RL in HPC batch job scheduling:

\begin{itemize}
    \item Can RL yield high-quality scheduling policy that is comparable to or even better than fine-tuned state-of-the-art scheduling policies, across various workloads and different optimization goals? 

    \item Is the RL-based scheduling policy only usable to its training workload or generally applicable to different workloads? 
    In another words, \rbt{will an RL-based policy still schedule jobs effectively on the new workloads that it never sees before}?

    \item What are the key factors that affect the learning efficiency of RL-based job schedulers?
\end{itemize}

We address these questions 
{in the design of RLScheduler}, a reinforcement learning based batch job scheduler.
Through extensive evaluations, we show that: first, with proper designs, RLScheduler is capable of learning high quality scheduling policy that is comparable to or even better than {the} state-of-the-art schedulers, on various (both synthetic and real-world) workloads or with vastly different optimization goals. Second, the model learned by RLScheduler works generally well even on job workloads that it never sees before, making it {sufficiently} stable to be used in practice. 

More importantly, {this study identifies} two key factors that
{affect the performance of RL-based batch job scheduler}:
1) the neural network structure of the agent; and 2) the variance of training datasets. To respond to these two factors, we propose a kernel-based deep neural network (DNN~\cite{GoodfellowIan2016DeepLearning}) and a trajectory filtering mechanism in RLScheduler. We believe that these two factors are general for other RL-based system-tuning problems and our solutions would provide useful insights for them too. 

{In summary, this study makes three key contributions:}
\begin{itemize}
    \item We build RLScheduler, the first reinforcement learning based batch job scheduler for HPC systems, to solve the adaption issue of existing batch job schedulers.
    
    \item We identify two key factors that affect the performance of reinforcement learning-based batch job scheduling and introduce corresponding solutions: kernel-based neural network and trajectory filtering mechanism to solve them.

    \item We conduct extensive evaluations to address the common concerns about utilizing RL in batch job scheduling. The results show the clear advantages of RLScheduler towards various workloads and changing system metrics.
\end{itemize}

The remainder of this paper is organized as follows: In \S\ref{sec:background} we introduce the necessary background about HPC batch job scheduling and deep reinforcement learning. In \S\ref{sec:discussion}, we discuss the challenges of applying deep reinforcement learning in batch job scheduling. In \S\ref{sec:design} we present the proposed RLScheduler and its key designs and optimizations. We present the main results (i.e. the RLScheduler and its performances) in \S\ref{sec:eval}, and compare with related work in \S\ref{sec:related}. We conclude this paper and discuss the future work in \S\ref{sec:conclude}.

	\section{Background}
	\label{sec:background}
	\subsection{HPC Batch Job Scheduling}
\label{subsec:back1}

This work discusses the job scheduling problem on HPC platforms, which offer homogeneous compute resources and host independent batch jobs. We discuss its key aspects briefly.

\subsubsection{Job Attributes}
On HPC platforms, a job presents several attributes, such as \textit{User ID}, \textit{Group ID}, \textit{Requested Processors}, and \textit{Submission Time}. Table~\ref{tab:attribute} summarizes some broadly seen job attributes. A more complete list of job attributes can be found in the Standard Workload Format (SWF)~\cite{Feitelson2014ExperienceArchive}.

For the job schedulers using \textit{priority functions}, selecting effective job attributes and fine-tune their combinations is a research topic, requesting manual efforts from domain experts or extensive research~\cite{tang2009fault,agarwal2006heuristics}. Comparatively, we build an RL-based scheduler \rbt{which simply takes all available job attributes and learns the most effective features automatically.}


\subsubsection{Workloads}

In the context of HPC batch job scheduling, workload usually includes a number of batch jobs and the timestamps addressing their submissions. A workload is typically characterized by the attributes of jobs and their arrival patterns. 
Due to the high variability and randomness of real-world workloads, it is hard to accurately model a workload. Researchers often use representative statistical values to characterize workloads, for example, the moments (e.g., mean, variance) of job runtime, job size, job arrival interval~\cite{lublin2003,Feitelson2001MetricsConvergence}.

HPC {workloads vary as new jobs submitted. But the variations may or may not change the workload characteristics. In this work, we consider workloads changes are those significant enough to vary the workload characteristics. For example, a load changes from short jobs to long jobs, or from small-scale jobs to large-scale jobs. As are expected, such changes can impact the system performance significantly and request the corresponding adaptions from job schedulers. We will describe more details about workloads characteristics in \S\ref{sec:eval}. }


\subsubsection{Scheduling Goal}
{The performance of job schedulers is measured by the optimization goals (or scheduling metrics).}
Different metrics address different {user expectations and lead to different scheduler designs accordingly.}  No single metric is considered as golden standard~\cite{Feitelson2001MetricsConvergence}. We summarize four widely used metrics/goals below.  

\begin{itemize}
    \item Minimize the \textbf{average waiting time (\textit{wait})}. It is the average time interval ($w_j$) between the submission and the start of a job.
    \item Minimize the \textbf{average response/turnaround time (\textit{resp})}. It is the average time interval between the submission time and the completion time of a job. This time is the waiting time ($w_j$) plus the job execution time ($e_j$).
    \item Minimize the \textbf{average bounded slowdown (\textit{bsld})}. Here, slowdown means the ratio of job turnaround time over its execution time $((w_j + e_j)/e_j)$, which overemphasizes short jobs with $e_j$ close to $0$. The bounded slowdown ($max((w_j + e_j)/max(e_j, 10), 1)$) measures job slowdown relative to given “interactive thresholds” (e.g., 10 seconds), which is considered more accurate.    
    \item Maximize \textbf{resource utilization (\textit{util})}, also called utilization rate, represents the average percentage of compute nodes allocated normalized by the entirety of nodes in the system over a given period of time. 
\end{itemize}

Previously, a scheduler is designed to optimize a fixed metric. For example, Shortest Job First (SJF), Smallest Job First, and F1---F4 in Carastan-Santos et al.~\cite{Carastan-Santos2017ObtainingLearning} target lowering average waiting time, increasing resource utilization, and minimizing average bounded slowdown, respectively. In the lifetime of a scheduler, when the system varies its scheduling metric, the system administrator will tune the scheduling policies manually. In this study, we leverage reinforcement learning and let the learning algorithm adjust its scheduling policies automatically for varying metrics. 

\subsubsection{Scheduling and Backfilling}
HPC platforms may provision multiple job queues and schedule jobs in different queues differently. Without loss of generality, batch jobs are usually submitted to \textit{batch queues} and scheduled by centralized job schedulers with backfiling techniques enabled~\cite{Mu2001UtilizationBackfilling}. 

The process is straightforward. In batch queues, when a job is selected, the system will seek for provisioning its requested resources. If success, the resources will be allocated and the job will start to run. Otherwise, the job will wait until its request is satisfied~\cite{yoo2003slurm}. In the mean time,  backfilling can be activated to search for the jobs whose resource allocations can be satisfied now without affecting the planned execution for the waiting job, to improve the efficiency of the system.

\subsection{Reinforcement Learning}
\label{subsec:drl}

\subsubsection{RL Concept}
Reinforcement learning (RL) is a group of machine learning techniques that enable agents to autonomously learn in an interactive environment by trials and errors~\cite{kaelbling1996reinforcement, Sutton2018ReinforcmentLearning}. {In this study},  we leverage {this} autonomy to build adaptive job schedulers for {varying workloads and goals}.

\begin{figure}[h!]
    \centering
    \includegraphics[width=0.9\linewidth]{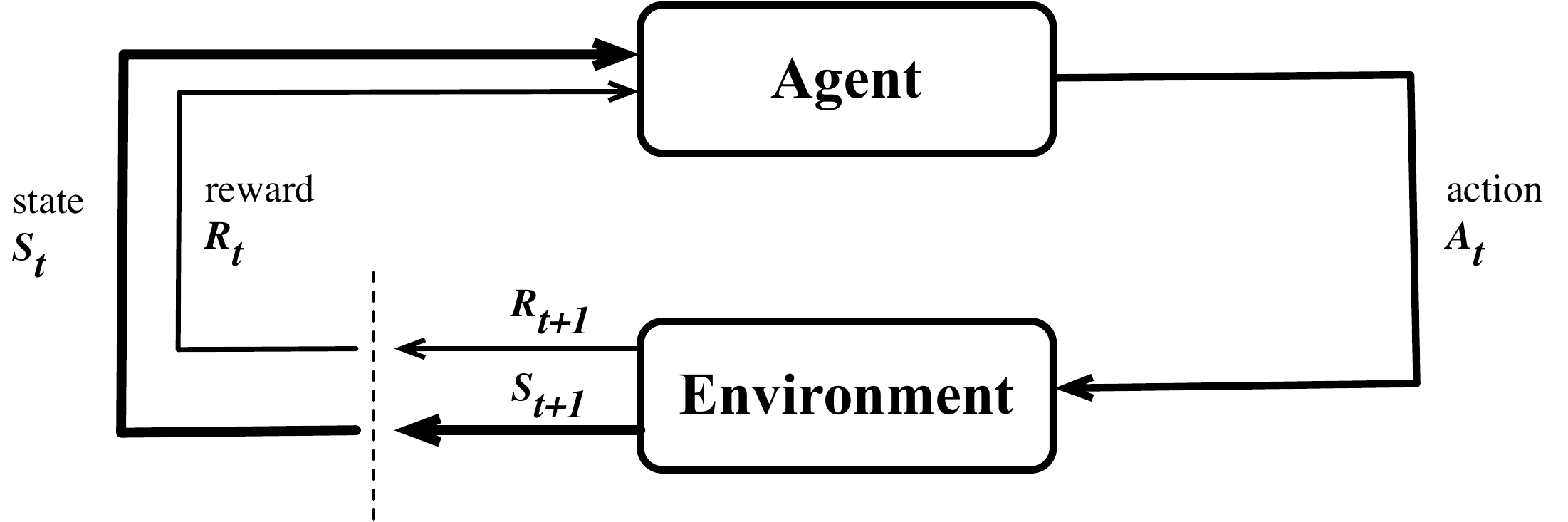}
    \caption{General framework of reinforcement learning.}
    \label{fig:rl}
\end{figure}

Fig.~\ref{fig:rl} shows a general RL framework. At each step $t$, the agent observes the corresponding state $S_t$, and takes an action $A_t$.  Consequently, the action will transfer the environment state from $S_t$ to $S_{t+1}$ and the agent will receive the reward $R_{t+1}$. In most cases, the agent does not have a prior knowledge on the environment state or the reward, and attain them gradually in the training process. The target of reinforcement learning is to  maximize the expected cumulative discounted reward collected from the environment.
The agent takes actions based on \textit{policies}, each defined as a probability of taking certain action at a given state. When the state space is enormous, memorizing all states becomes infeasible.  Deep Neural Network (DNN) can be used to estimate the probability. The reinforcement learning using DNN to model the policy is called Deep Reinforcement Learning (DRL).

\subsubsection{RL training methods}
Reinforcement learning has a large number of training methods, 
classified in different ways~\cite{sutton2018reinforcement}. But, a key difference among them is the training strategies, i.e., what the RL agent learns. The \textit{policy-based RL} directly learns the policy, which will output an action given a state; and \textit{policy gradient} method is a typical example of them~\cite{sutton2000policy}. The \textit{value-based RL} learns proper value of each state, which can indirectly output an action by guiding the agent to move towards better state; and \textit{Q-learning} method is a typical example~\cite{watkins1992q}. Between these two methods, \textit{policy gradient} is proven to have strong convergence guarantees~\cite{sutton2000policy} and become our first choice. This is mostly due to the high variance of batch job scheduling, which may lead to oscillations in \textit{Q-learning}. To alleviate the known performance issues of \textit{policy gradient}, we follow the Actor-Critic model~\cite{konda2000actor} in RLScheduler to combine both policy-based and value-based learning for better training efficiency. {We return to this in \S~\ref{sec:design}}.

    \section{Discussion on Challenges}
    \label{sec:discussion}
    {At first glance, scheduling batch jobs with a deep reinforcement learning agent seems intuitive by repeating three simple steps: 1) take the waiting jobs and idle compute resources of the target HPC environment as the input for a deep neural network (DNN); 2) use the DDN as the current scheduling policy to select a `best` job as the action; 3) apply the action back to the environment. The training process repeats the three steps until the last job in the job sequence is scheduled, which creates one sampled \textit{trajectory}, and then computes the reward based on a given metric.
} With {sufficient} trajectories and their rewards, the \textit{policy gradient} algorithm can be used to update the policy (DNN) to maximize expected {rewards} of these trajectories, {indicating} a better scheduling algorithm. 

Although the {process}
is 
standard for all \textit{policy gradient} RL, {the techniques used in}
each step is specific to {the target problem and can affect the training efficiency and the agent correctness significantly. In this study, we address} 
two key challenges in the RL-based batch job scheduling problem.

\subsubsection{RL Network Architecture}
\begin{figure}[h!]
    \centering
    \includegraphics[width=\linewidth]{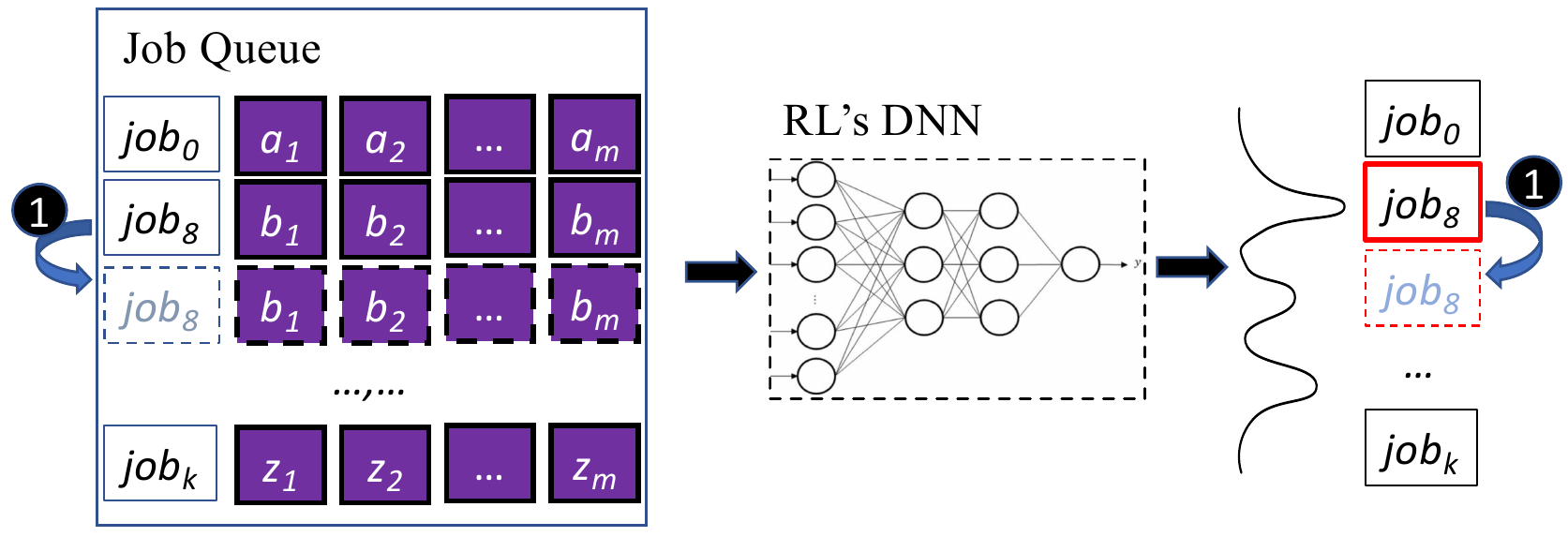}
    \caption{The DNN-based RL agent reads waiting jobs and selects a job as the scheduling decision.}
    \label{fig:jobs}
\end{figure}

In Fig.~\ref{fig:jobs}, we show how an DNN-based RL agent makes scheduling decisions: it takes the waiting jobs and their features (e.g., $a_{1 \to m}$) as input vector and outputs a probability distribution of each job being scheduled next. The {job with the highest probability ($job_8$ in this example) should be the selected job}. One key issue here, however, is {the job orders in the waiting queue could change easily.} As `step 1' shown in the figure, $job_8$ may {hold a different position} in the queue next time, for example, from the second to the third. But, {the RL's DNN should still select $job_8$ as the best even its placement is different.}

In general, there are two ways to achieve this. One way is to consider the job orders in queue as the translation or deformation of the inputs  and  learn these deformation by feeding  the  DNN  with  more training data. The other option is to make the DNN insensitive to job orders, and assign a job with the same probability regardless its orders in the job queue. The former approach looks intuitive. Nevertheless, it usually requests much more training data, takes longer time to converge, and may deliver lower model accuracy.  In this study, we take the latter approach  and  design  a  new  kernel-based DNN architecture to be insensitive to job orders. We experimented on both approaches and confirmed that ours obtains much better performance in model training efficiency and accuracy (\S\ref{sec:eval}).


\subsubsection{High Variance in Samples}
The \textit{policy gradient} algorithm essentially is a Monte-Carlo method, which samples a large number of \textit{trajectories} and uses their results to adjust the DNN (representing current policy). The key for this to work is that these samples can accurately reflect the quality of policy, which, however, might not be true in batch job scheduling. For example, if a sampled job sequence arrive sparsely, then each job can be instantly scheduled to run, their \textit{job waiting times} will be $0$ no matter what scheduling policy is used. On the contrary, if the sampled jobs arrive {at the same time}, their \textit{waiting times} will be {relatively long} no matter what scheduling policy is used. If samples a lot of the first cases, then RL agent may misinterpret the scheduling policy as good, which leads to unstable or even non-converged results.

\begin{figure}[h!]
    \centering
    \includegraphics[width=0.8\linewidth]{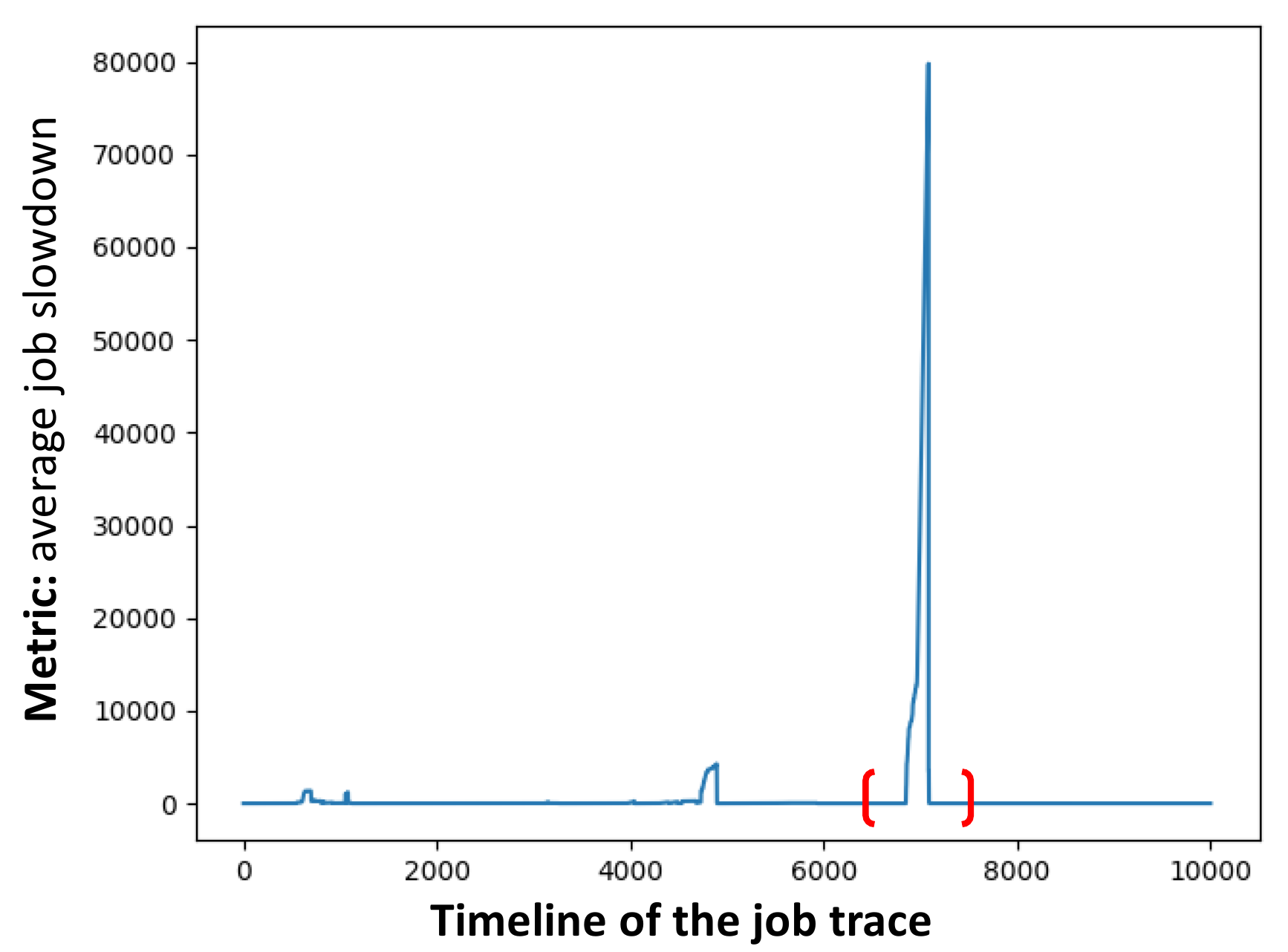}
    \caption{The \textit{average bounded slowdown} of scheduling sequence of 256 jobs in PIK-IPLEX-2009 job trace.}
    \label{fig:variance}
\end{figure}

The key is how high the variance of the batch job scheduling can be in real world. In Fig.~\ref{fig:variance}, we show an example of using the SJF (Shortest Job First, \rbt{i.e., always selecting the job with the shortest requested runtime}) scheduling algorithm to schedule a sequence of 256 jobs sampled from the PIK-IPLEX-2009 job trace, which is a real-world job trace collected from IBM iDataPlex Cluster~\cite{Feitelson2014ExperienceArchive}. Here, the \rbt{\textit{vertical} axis} shows the average job slowdown calculated from scheduling the whole sequence, and the \rbt{\textit{horizontal} axis} shows the timeline of the job trace (we show the first 10K jobs as an example).

From this figure, we can see that, in most of the time, the job slowdown is close to $1$, which indicates the jobs barely wait in the queue. But there are short period of time (e.g., the red range) where the average job slowdown reaches 80K, which indicates a really long job waiting time. The variance is so high that it has two negative impacts. First, one `bad` trajectory will diminish what RL agent has learned as we have discussed. Second, too many `good' trajectories will barely teach RL agent anything during training, because no matter what scheduling policy it currently holds, the slowdown is gonna be $1$. In RLScheduler, we propose strategies to eliminate the high variance and ensure the convergence of RL training even facing a workload like PIK-IPLEX-2009. More details will be discussed in the next section.

	\section{Design and Implementation} 
	\label{sec:design}
	RLScheduler uses reinforcement learning to derive adaptive policies for scheduling HPC batch jobs towards the varying workloads and optimization goals. Our approach is fundamentally different to the previous job schedulers, which rely on the expert knowledge about workloads, job attributes, and the optimization goals (discussed in \S\ref{subsec:back1},~\S\ref{subsec:drl}). RLScheduler is independent \rbt{from} the knowledge and effort from experts. The only inputs it takes are the job traces and optimization goals, then it learns the scheduling by itself.

This section first overviews the RLScheduler's design and implementation, then discusses its two key techniques: kernel-based neural network and variance reduction methods. 

\subsection{RLScheduler Overview}

\begin{figure}[h!]
\centering
\includegraphics[width=\linewidth]{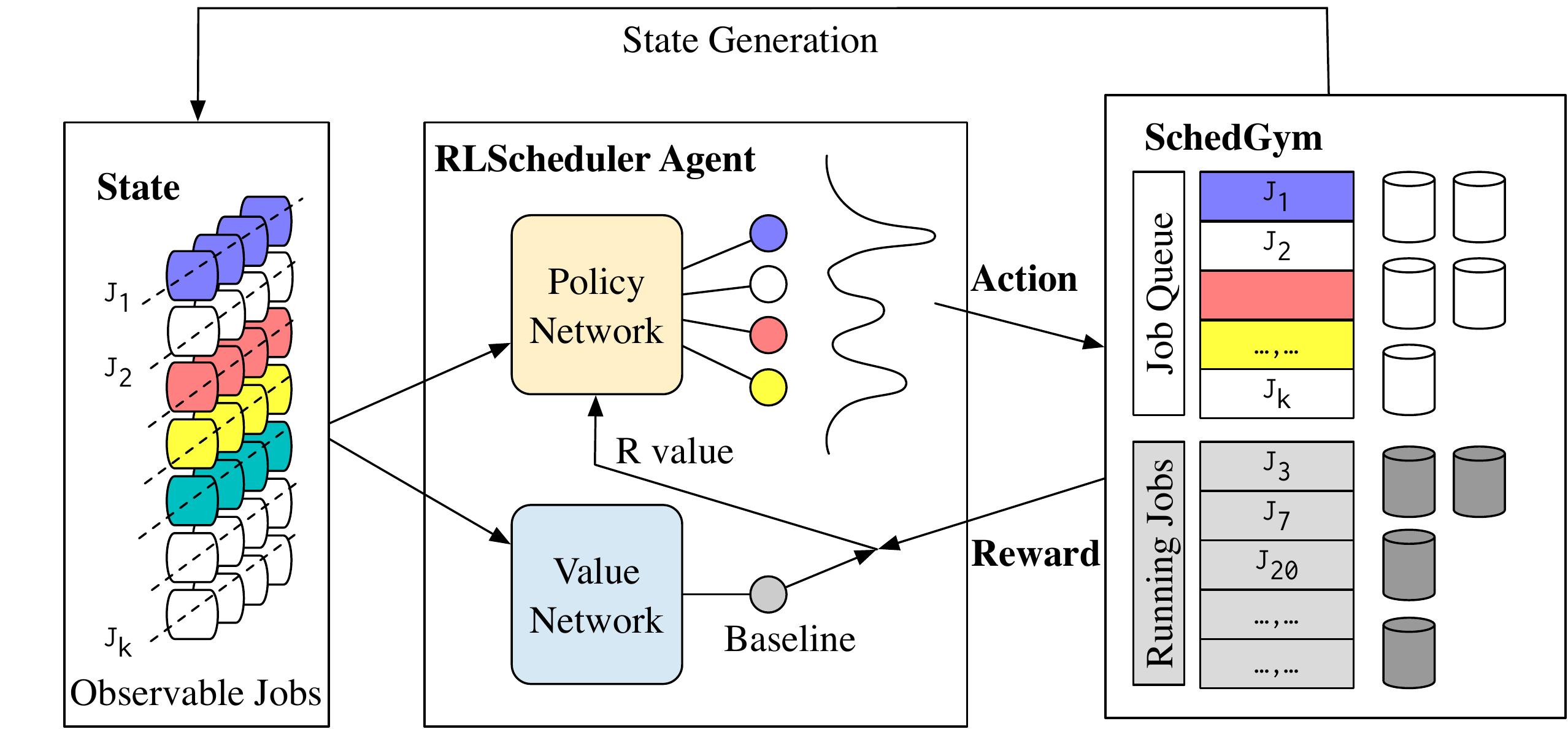}
\caption{The overall architecture of RLScheduler.}
\label{fig:overall}
\end{figure}

Fig.~\ref{fig:overall} shows the RLScheduler architecture and its three major components: the \textit{Agent}, the job scheduling \textit{Environment}, and the environment \textit{State}. 
In each step of the training process, the RLScheduler agent observes a state and takes an action. The state is collected and built from the environment. The action will be applied to the environment and consequently generate the next state and a reward. Across steps, the agent learns from its actions and the associated rewards. 

In particular, a reward is the feedback of the environment on an agent's action. It serves as the key to guide the RL agent towards the better policies. In RLScheduler, reward is a function addressing a user-given optimization goal. For instance, if the optimization goal is to minimize \textit{average bounded slowdown} ($bsld$), the reward can simply be $reward=-bsld$, which means the RL will maximize the $reward$ by minimizing the average bounded slowdown. If the optimization goal is to maximize \textit{resource utilization} ($util$), the reward can be $reward=util$, which will directly maximize the utilization.

Note that, the reward is supposed to be collected in each learning step whenever the agent takes an action. However, for most of the scheduling metrics, such as \textit{average waiting time} or \textit{average bounded slowdown}, the calculation can not be done until the whole job sequence gets scheduled. 
Thus, in the middle of scheduling a job sequence, we just return rewards $0$ to each action and calculate the accurate reward for the entire sequence at the last action. This does not affect RL training as only the accumulated rewards are used for training.  


\subsection{{RLScheduler Kernel-based Neural Network}}
\label{subsec:agent}

The RL agent's deep neural networks play the key role in learning. In RLScheduler, we leverage two networks: \textit{policy network} and \textit{value network} following the actor-critic model, to conduct the learning. They take the roles of generating scheduling actions and facilitating the training respectively.

\subsubsection{Policy network}
The policy network takes the updated environmental state as its input and directly outputs an action determining which job to run next. 

The key here is to make policy network insensitive to the order of jobs (discussed in \S\ref{sec:discussion}). 
\begin{figure}[h!]
  \centering
  \includegraphics[width=0.9\linewidth]{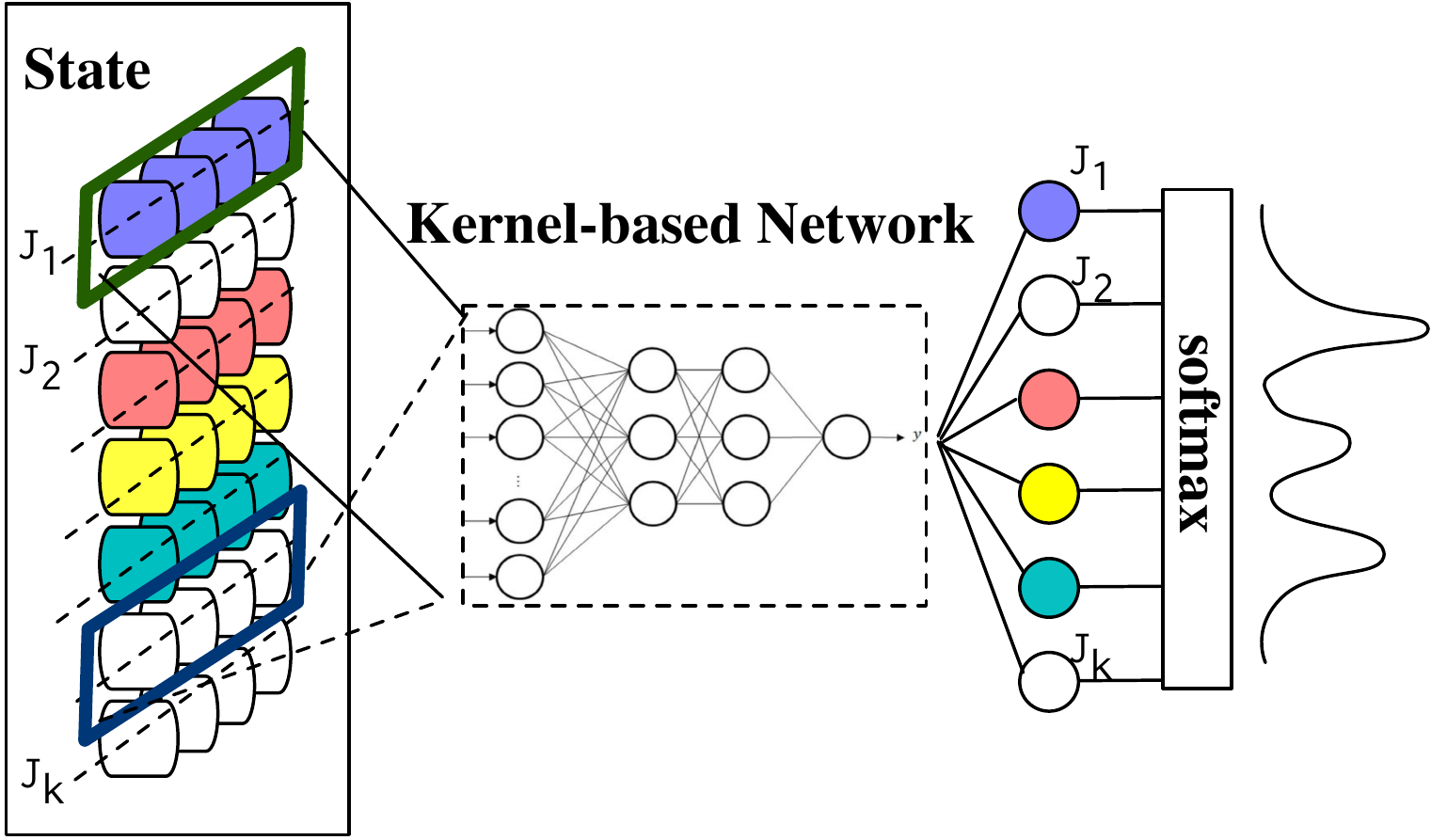}
  \caption{The RLScheduler policy network structure. Its core is a kernel-based neural network.}
  \label{fig:nn}
\end{figure}
To this end, we design a kernel-based network as the policy network. Fig.~\ref{fig:nn} shows the network in detail. \rbt{The kernel network itself is a 3-layer fully connected network, structured the same as 3-layer perceptron (MLP)~\cite{bourlard1988auto}. The difference is the kernel-based network will be applied onto} each waiting job one by one like a window. For each waiting job, the network outputs a value, a calculated `score' of the job. The values of all waiting jobs form a vector. We then run softmax on the vector to generate a probability distribution for each waiting job. In this way, once jobs are reordered, their probabilities will also be reordered accordingly. 
This design is inspired by the kernel function in convolutional neural networks (CNN)~\cite{krizhevsky2012imagenet}. But, we eliminate the pooling and fully connected layers of CNN as they are sensitive to job orders. We later confirmed the advantage of our proposal in the evaluation section by comparing it with CNN and other MLP networks. 

The probability distribution of each waiting job serves two purposes: 1) during training, it is sampled to obtain the next action. Sampling enables us to keep exploring new actions and policies; 2) during testing, it is directly used to select job with the highest probability to ensure the best decision. There is no exploration anymore. 

The kernel-based design makes the policy network relatively simple. Together with the small input dimension (i.e., a single job's attributes each time), the parameter size of policy network becomes extremely small.
In RLScheduler, we are able to control the parameter size of the policy network less than 1,000, which \rbt{consequently} improve the training efficiency.

\subsubsection{Value network}
RLScheduler also includes a \textit{value network} to formulate an Actor-Critic model to improve training efficiency. It takes a entire job sequence as inputs and outputs a value to indicate the expected reward of that sequence. 

\begin{figure}[h]
\centering
\includegraphics[width=0.7\linewidth]{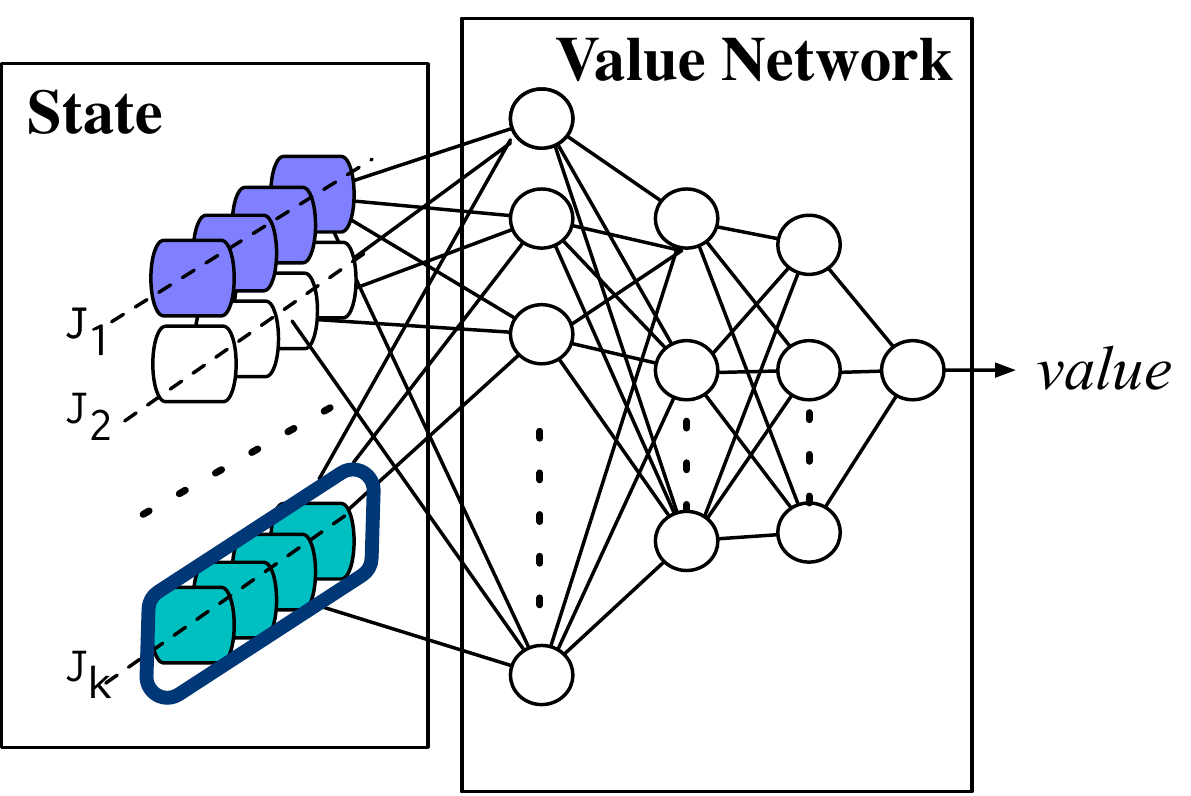}
\caption{The RLScheduler value network structure. Its core is a 3-layer multiple layer perceptron network (MLP).}
\label{fig:value}
\end{figure}

Fig.~\ref{fig:value} shows the \textit{value network} internal, which is a 3-layer MLP network, but does not have the kernel mechanism. To work with MLP, the vectors of all jobs will be concated and flatten before feeding into the network. 

The \textit{value network} is trained along with the \textit{policy network}. Specifically, for a sequence of jobs, after the \textit{policy network} makes all the scheduling decisions, we collect the rewards, then use this to train the \textit{value network} to predict the reward of a given job sequence. 

The output of \textit{value network} can be intuitively considered as the expected reward ($exp_r$) of a set of jobs based on the agent's current policies. It indicates how best the agent can do on this set of jobs historically. 
When we train the \textit{policy network}, instead of directly using the accumulated rewards ($r$) collected from the environment, we can use ($r - exp_r$) to train the policy. This difference can be intuitively considered as the improvement of current policy over historical policies on this set of jobs. 
This strategy helps reduce the variance of inputs and lead to better training efficiency.


\subsubsection{The Inputs of DNNs}
\label{subsec:state}

The inputs of both policy and value networks are the \textit{state}, which includes both job attributes and available resources in the systems. RLScheduler uses a vector $v_{j}$ to embed such state info into each job ($j$). Then, multiple jobs will form a matrix as shown in Fig.~\ref{fig:nn}. 

Each vector first contains all the available attributes of a job, such as job arriving time and request processors. A full list is shown in Table~\ref{tab:attribute}. %
In addition to job attributes, the vector also contains \textit{available resources} to indicate the available resource in the system. The priority of a job actually varies depending on the currently available resources.

One practical issue of using DNNs to read all waiting jobs is the number of waiting jobs changes, but our DNNs only take fixed-size vector as inputs. To solve this, we limit RLScheduler to only observe a fixed number (\verb|MAX_OBSV_SIZE|) of jobs. If there are fewer jobs, we pad the vector with all $0$s job vector; if there are more jobs, we cut-off them selectively. The number of observable jobs is a configurable training parameter. We set it to 128 in RLScheduler by default, as many HPC job management systems, such as Slurm, also limit the number of pending jobs to the same order of magnitude~\cite{slurmweb}.
When cut-off extra jobs, we simply leverage FCFS (first come first serve) scheduling algorithm to sort all the pending jobs and select the top \verb|MAX_OBSV_SIZE| jobs. 


\subsection{RLScheduler Variance Reduction}
\label{subsec:reduction}
As discussed in \S\ref{sec:discussion}, the high variance in HPC batch jobs will impose significant challenges onto reinforcement learning. When we randomly sample job sequences from real-world job trace for training, such as the PIK-IPLEX-2009, RL agent will experience the `easy sequences' which mean any scheduling algorithm will lead to good results and the `hard sequences' which mean any scheduling algorithm will lead to bad results. From the view of RL training, both cases are destructive: the `easy sequences' do not provide any meaningful knowledge to the agent; while the `hard sequences' simply confuse it. 

Recent studies have seen similar issues for training RL in such `input-driven' environments and proposed to reduce the variances by memorizing the scheduling results for the same job-arrival sequence and let the RL learn from the relative improvement on its own rather than from the absolute reward values~\cite{Mao2018VarianceEnvironments,Mao2018LearningClusters}. This solution has two major drawbacks in batch job scheduling: 1) it does not solve the issue of `easy samples', which could take a large portion of a real-world job trace and do not help in training RL except taking computation time; 2) memorizing history only helps when there are repeated visits on the same job sequence. Given the size of real-world job traces, for example the PIK-IPLEX-2009 has over 700K jobs, the re-visits are expected to be so rare that the memorized history is sparse and insufficient to use.

Instead, we introduce \textit{trajectory filtering} in RLScheduler. The key idea is to filter out some job sequences during training, so that the RL agent will see sequences with controlled variances and learn in a more stable way. In particular, it filters the `easy sequences' out since they will not contribute info to improve the RL agent. For the `non-easy sequences', it categorizes all sequences into two ranges and trains the RL agent in two steps. The first step contains job sequences whose variances fall into a specific range (\textbf{$R$}). So that the agent can learn in a more stable way and converge faster. The second step trains on all the job sequences. Although they still have high variances, since the RL agent has already converged, our experiences show that the agent is hard to be misled again. 

\begin{figure}[h!]
    \centering
    \includegraphics[width=0.9\linewidth]{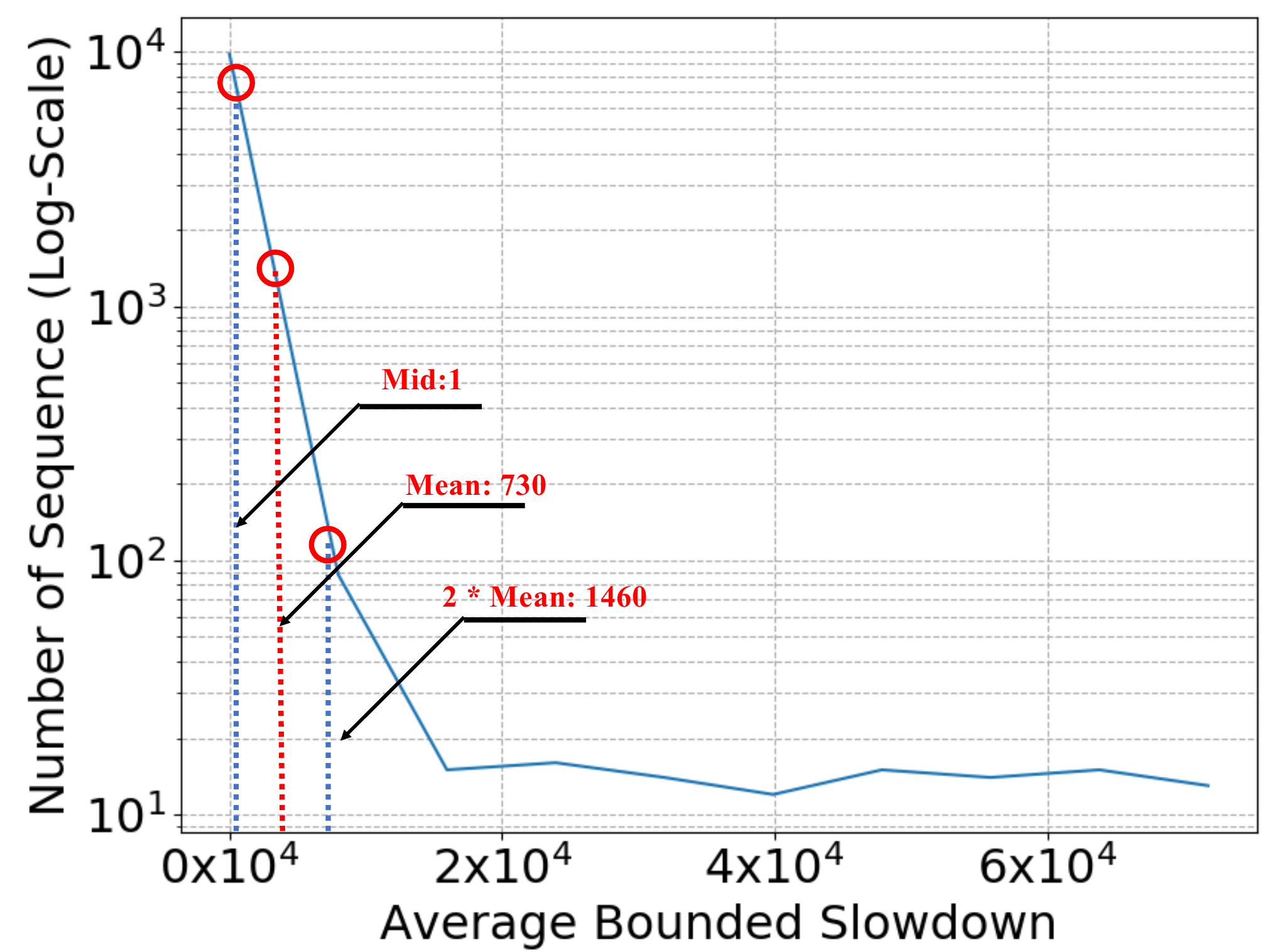}
    \caption{The distribution of the \textit{average bounded slowdown} of scheduling sequence of 256 jobs in PIK-IPLEX-2009 job trace.}
    \label{exp:slowdown-pik}
\end{figure}

Then, finding a good range ($R$) to rule out high variant job sequences becomes important. To do so, we use a known heuristic scheduling algorithm, i.e., Shortest Job First (SJF), to schedule a number of randomly sampled job sequences from the job trace and collect their metrics values first. We then calculate its key statistical values: \textit{median}, \textit{mean}, and \textit{skewness}. For a high variant job trace, such as PIK-IPLEX-2009 trace, the resulting distribution will look like Fig.~\ref{exp:slowdown-pik}. Here, {\it x-axis} is the metrics value, i.e., the average bounded slowdown; {\it y-axis} is the number of sampled sequences whose scheduled metrics equal to that value. We also plot its \textit{median}, \textit{mean}, and \textit{2*mean} in the figure. 
Based on this distribution, we simply determine the metrics range $R$ of the first training step as $R=\left(median, 2*mean\right)$. In this way, we remove the `easy sequences` which take half of the samples ($median$), and also control the variance within doubled $mean$, which is a much smaller number in the highly skewed distribution.


Note that, not all job traces need \textit{trajectory filtering}. Such as SDSC-SP2, more stable job traces can be directly trained. The scheduling metrics is also an important factor. Some metrics, such as resource utilization typically has much smaller variances, hence do not improve much with trajectory filtering.

\subsection{{SchedGym Environment}}
\label{subsec:env}

Training RLScheduler in a running HPC platform is impractical since reinforcement learning requires an enormous number of interactions with the environment to learn.
In RLScheduler, we implement {SchedGym} as the simulated environment to interact with the RL agents.

SchedGym is based on OpenAI Gym toolkit~\cite{brockman2016openai}. It takes a standard SWF~\cite{Feitelson2014ExperienceArchive} job trace as input and simulates how HPC computation platform works.
Starting from an idle cluster, it loads jobs from job trace one by one. When new job arrives or running job finishes, 
SchedGym will query the scheduler and act based on the returned action. When the available resources are insufficient to host the scheduled job, SchedGym can backfill possible jobs whose executions will not \rbt{impact} the jobs under scheduling. 
The actual runtime of a job is retrieved from the SWF job traces. Since we target homogeneous HPC in this study, we assume the runtime will not change. Note that, the accurate runtime will not be available to the schedulers, instead, only the \textit{requested runtime} is available to schedulers.

	
	\section{Evaluation}
	\label{sec:eval} 
	\subsection{Evaluation Setup}

We implement RLScheduler based on the Proximal Policy Optimization (PPO) algorithm from OpenAI Spinning Up~\cite{schulman2017proximal} using Tensorflow~\cite{Abadi2016TensorFlow:Learning}.

The training in RLScheduler works in epoch. In each epoch, it samples multiple trajectories from the environment. Each trajectories includes a series of interactions between the agent and the environment. The collected rewards will be used to update the agent. In RLScheduler, we take $100$ trajectories in each epoch and each trajectory contains the scheduling decisions of $256$ continuous jobs. In-between epoch, RLScheduler runs $80$ iterations to update its policy network and value network separately. The learning rate is $10^{-3}$.
More hyper-parameters can be found in the source code\footnote{\url{https://github.com/DIR-LAB/deep-batch-scheduler}}~\cite{rlscheduler}.


{We list the job traces used in the evaluations in Table~\ref{tab:jobtrace}, and categorize them into two groups. The first group addresses}
the real-world traces from SWF archive~\cite{Feitelson2014ExperienceArchive}. {The second group addresses} the synthetic traces generated based on a widely used workload model proposed in \cite{lublin2003}. We used different parameters in the model and generated two traces with different characteristics. 
As the sizes of these job traces are largely different, we leveraged the first 10K jobs from them in our evaluations.
\begin{table}[ht]
	\small
	\caption{List of job traces}
	\begin{center}
		\label{tab:jobtrace}
		\begin{tabular}{cccccc}
			\toprule
			Name & Date &$size$ & $i_t$(sec) & $r_t$(sec) & $n_t$  \\		
			\midrule
			SDSC-SP2 & 1998  & 128 & 1055 & 6687 & 11 \\
            HPC2N   & 2002 & 240 & 538 & 17024 & 6 \\
			PIK-IPLEX & 2009 & 2560 & 140 & 30889 & 12 \\
			ANL Intrepid & 2009 & 163840 & 301 & 5176 & 5063 \\
			\midrule
			Lublin-1 & -  & 256 & 771 & 4862 & 22 \\
			Lublin-2 & -  & 256 & 460 & 1695 & 39 \\
			\bottomrule
		\end{tabular}
	\end{center}
\end{table}

The key characteristics of these traces are also shown in the table, including the total number of processors in the cluster ($size$), average job arrival interval ($i_t$), average requested runtime ($r_t$), and average requested processors ($n_t$). In summary, the job traces are quite diverse in the presented characteristics.  

\begin{table}[ht]
	\caption{List of schedulers}
	\label{tab:funcs}
	\begin{center}
		\begin{tabular}{cl}
			\toprule
			Name & priority function \\
			\midrule
			FCFS & $score(t) = s_t$\\
			SJF & $score(t) = r_t$\\
			WFP3 & $score(t) = -(w_t/r_t)^3*n_t$\\
			UNICEP & $score(t) = -w_t/(log_2{(n_t)}*r_t) $\\
			\midrule
			F1 & $score(t) = log_{10}(r_t)*n_t + 870*log_{10}(s_t)$\\
			\bottomrule
		\end{tabular}
	\end{center}
\end{table}

To evaluate RLScheduler, we compare with existing \textit{priority function}-based schedulers, including several heuristic schedulers and a state-of-the-art learning-based scheduler.
Table~\ref{tab:funcs} reports the priority functions used in these schedulers.
Here, FCFS schedules jobs in the same order as they were submitted (i.e., using $s_t$). 
SJF schedules jobs based on how long the job will run (i.e., using $r_t$). WFP3 and UNICEP~\cite{tang2009fault} belong to the  scheduler family that combines multiple factors. More specifically, they favor jobs that have shorter runtime, request fewer resources, and experience longer waiting time, representing the expert knowledge in tweaking the priority functions. Scheduler F1 is the best scheduler selected from~\cite{Carastan-Santos2017ObtainingLearning}. It was built on brute force simulation and non-linear regression, and represents the state-of-the-art batch job scheduler for the goal of minimizing \textit{average bounded slowdown}.

In the following subsections, we report the evaluation results of RLScheduler under various scenarios. The results mainly address the following questions about RLScheduler:
\begin{itemize}
    \item {Whether the new designs (\textit{kernel-based neural network and trajectory filtering mechanism}) improve the training performance of RLScheduler?}
    \item {How well are RLScheduler's training and performance towards: \textit{different HPC workloads}, \textit{different scheduling metrics}, or even \textit{combined scheduling metrics}?}
    \item {Will a scheduling policy that RLScheduler learns still be applicable to an unseen, new workload?}
    \item What is the computational overhead of RLScheduler? 
\end{itemize}



\subsection{RLScheduler Design Evaluations}
This section examines the key designs of RLScheduler. In particular, we measure the performance improvement of the kernel-based neural network and trajectory filtering mechanism during RLScheduler training.

\subsubsection{Kernel-based Neural Network Performance}

\begin{table}[ht]
    \begin{center}
	\caption{\rbt{The network configurations of different policy network designs, including our design (RLScheduler)}}
	\label{tab:nncompare}
	\begin{tabular}{ccc}
		\toprule
		Name & Layers & Size of each layer\\
		\midrule
		MLP\_v1& 3 & $128,128,128$\\
		MLP\_v2& 3 & $32,16,8$\\
		MLP\_v3& 5 & $32,32,32,32,32$\\
		\midrule
		LeNet~\cite{lecun1998gradient} & 6 & 2x(conv2d, maxpooling2d), dense\\
		\midrule
		RLScheduler & 3 & $32,16,8$\\
		\bottomrule
	\end{tabular}
	\end{center}
\end{table}

\begin{figure}[h]
	\centering
	\includegraphics[width=0.8\linewidth]{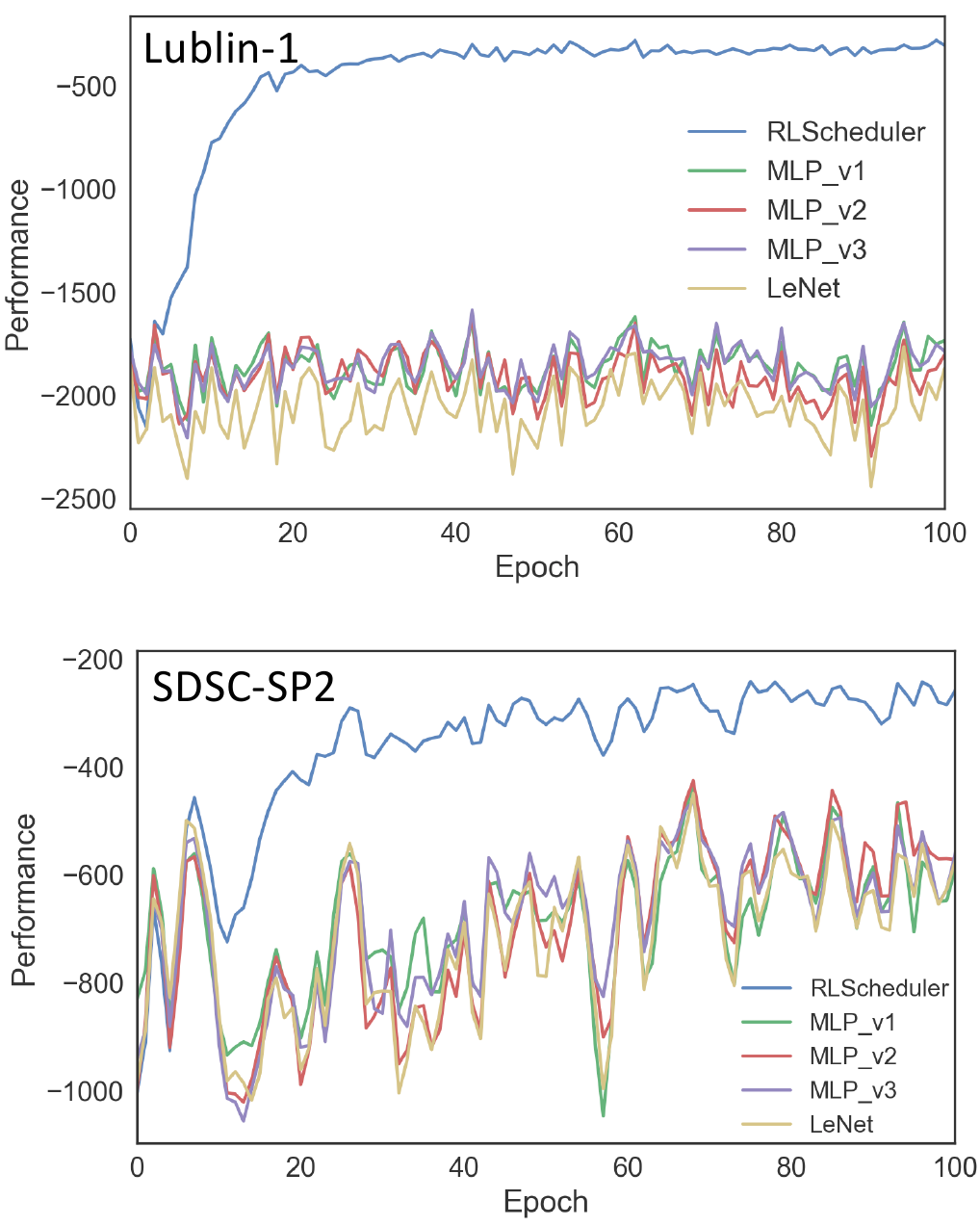}
	\caption{The training efficiency of different RLScheduler policy networks on two job traces (Lublin-1 and SDSC-SP2). \textit{The \rbt{horizontal} axis shows the total number of training epoch; the \rbt{vertical} axis shows the performance of the agent, referring to $- bsld_{avg}$. The larger \rbt{vertical} axis value indicates a smaller average bounded job slowdown and is better.}}
	\label{fig:training}
\end{figure}

To measure kernel-based neural network, we \rbt{compared the training efficiency of RLScheduler using different policy neural} networks, including \rbt{convolution neural network (CNN) and multiple layer perceptron (MLP) networks. We selected them because they are broadly used in similar reinforcement learning-based system optimization studies~\cite{Mao2016ResourceLearning,Mao2018LearningClusters}}. For MLP, we evaluated three settings. For CNN, we used the standard LeNet~\cite{lecun1998gradient}. The parameters of these networks are listed in Table~\ref{tab:nncompare}. 
\rbt{We conducted all the experiments under the same setting: targeting the same metrics (average bounded job slowdown); sharing the same value network and hyper-parameters except the policy network}. Fig.~\ref{fig:training} presents the \rbt{training curves} of different policy neural networks on two job traces (Lublin-1 and SDSC-SP2). 

{In summary,} the results support two points. First, RLScheduler converges fast and reaches a good \rbt{performance} within 20 epoch, suggesting high efficiency of the kernel-based network and the policy gradient reinforcement learning algorithm. Second, RLScheduler \rbt{with kernel-based policy network} converges \rbt{much faster than} other networks. We evaluated MLP with different configurations and observed negligible difference. LeNet recognizes jobs with the similar kernel like our solution, but delivers much worse performance than ours. It suggests that, the later pooling and fully connected layers in LeNet mix the job orders and \rbt{degrade the training efficiency}. 


\subsubsection{Trajectory Filtering Performance}

\begin{figure}[h]
	\centering
	\includegraphics[width=\linewidth]{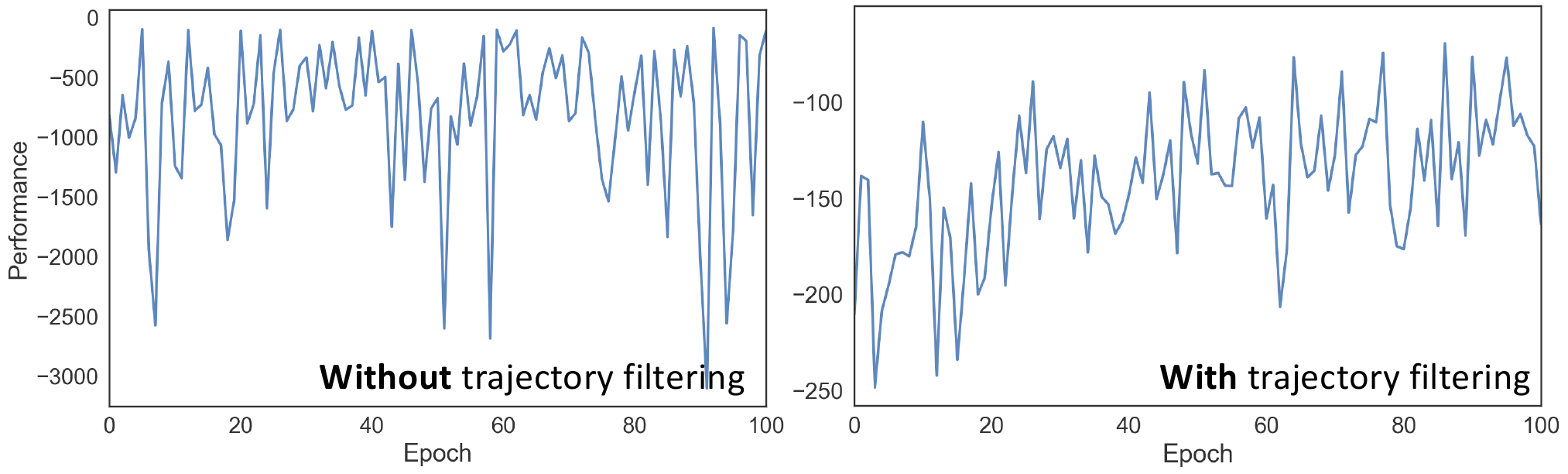}
	\caption{The training curves of RLScheduler on PIK-IPLEX-2009 job trace \textit{with} and \textit{without} trajectory filtering.}
	\label{exp:pik}
\end{figure}

\rbt{To measure the trajectory filtering mechanism, we compared the training efficiency of RLScheduler} on job trace PIK-IPLEX-2009 \textit{with} and \textit{without} trajectory filtering. \rbt{Fig.~\ref{exp:pik}} reports the results. It shows that, without trajectory filtering, the training does not converge even after 100 epoch; with trajectory filtering enabled (the $R$ range is $\left(1, 1460\right)$ as suggested in Fig.~\ref{exp:slowdown-pik}), RLScheduler converges. When taking a closer look at the job trace, we find that this job trace has extremely high variance under the metrics of \textit{average bounded slowdown}. Not surprisingly, without trajectory filtering the RLScheduler agent gets confused by the rare `hard sequences'. Filtering out the sequences that have huge {job slowdown} hence significantly improves the training stabilization and efficiency.

\subsection{RLScheduler Performance on Various Job Traces}

This section reports RLScheduler performance on different job traces under the scheduling metric of \textit{average bounded slowdown}. We will discuss RLScheduler performance on different metrics in a separate section later. 

\subsubsection{Training on Different Workloads}
\begin{figure}[h!]
	\centering
	\includegraphics[width=\linewidth]{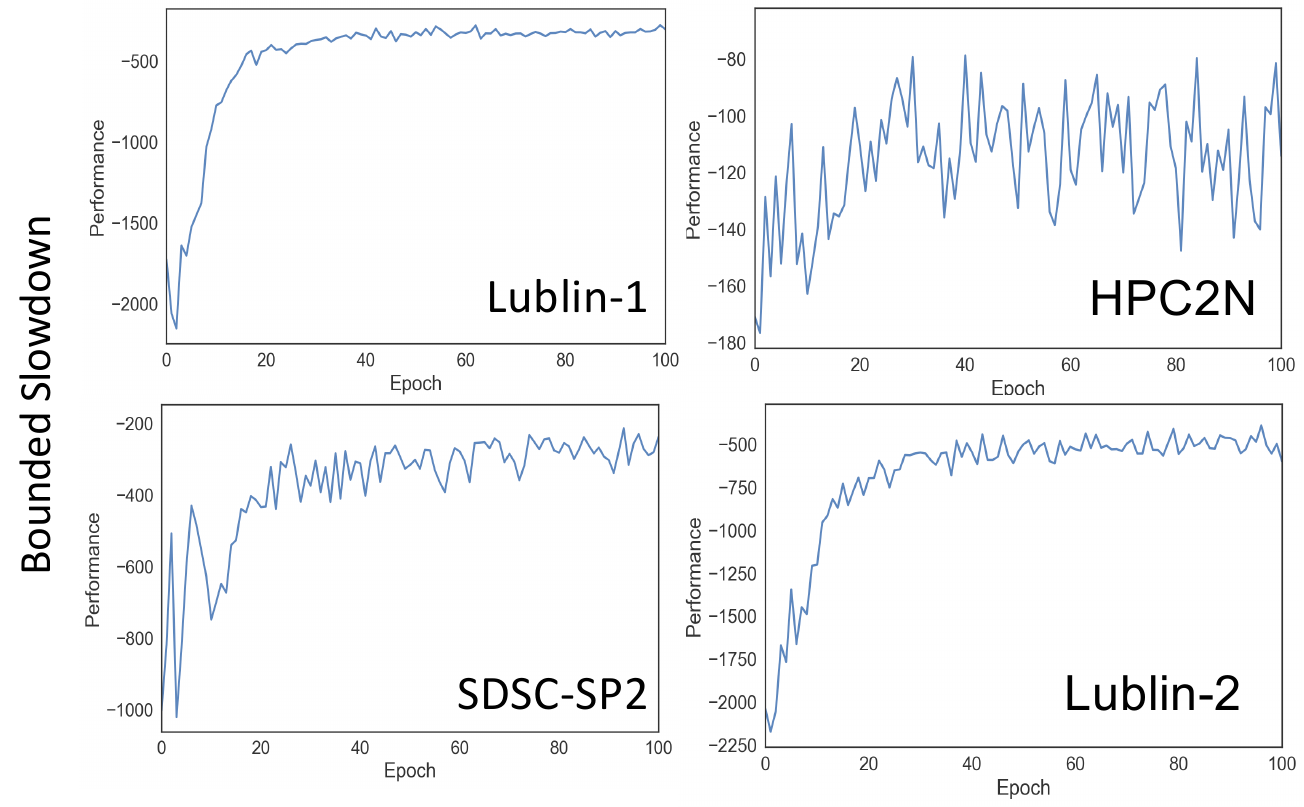}
	\caption{The training curves of RLScheduler on four different workloads. \textit{Here, all \rbt{vertical} axis are the average bounded slowdown calculated after scheduling the whole epoch. The \rbt{horizontal} axis shows the training epoch.}}
	\label{exp:train-workloads}
\end{figure}

Fig.~\ref{exp:train-workloads} reports the training curves of RLScheduler on different job traces, including two real-world workloads (HPC2N and SDSC-SP2) and two synthetic workloads (Lublin-1 and Lublin-2).

The results show RLScheduler converges in all of the workloads within 100 training epoch, again showing its learning efficiency. Interestingly, we observe different convergence patterns across workloads. As discussed in \S\ref{subsec:reduction}, this is \rbt{largely} due to the different variances of the job traces themselves. And because of this, for some job traces, it is necessary to place the \textit{trajectory filtering} to guarantee the convergence, as shown in the previous subsection.

\subsubsection{Scheduling on Different Workloads}
Next we show the performance of RLScheduler when it actually schedules the workloads. We repeated the evaluations 10 times and reported the average. In each experiment, we scheduled a random job sequence that contains long continuous jobs (1,024) from the corresponding workloads. Note that, we selected much longer job sequences (1024) for testing than the job sequences (256) used for training. This helps detect whether RLScheduler over-fits the training datasets. Also, across different scheduling algorithms, we used the same $10$ random job sequences to make fair comparisons. Table~\ref{tab:workloads} presents the results, in which the best scheduler for each job trace is marked bold. Here, we show two set of scheduling results: with or without backfilling. As an independent feature of batch job scheduler, backfilling can be enabled/disabled on any job scheduler and may have significant impact on the performance. Hence, we exam RLScheduler in both cases.

\begin{table}[ht]
    \caption{Results of scheduling different job traces.}
    \label{tab:workloads}
    \begin{center}
	\begin{tabular}{|c|c|c|c|c|c|c|}
		\hline
		\textit{Trace}     & FCFS   & WFP3    & UNI    & SJF    & F1     & \textbf{RL}     \\ \toprule
		\hline
		\multicolumn{7}{|c|}{\textit{Scheduling \textbf{without} Backfilling}}
		\\\hline
		\textit{Lublin-1} & 7273.8  & 19754   & 22275   & 277.35  & 258.37           & \textbf{254.67}  \\ \hline
		\textit{SDSC-SP2} & 1727.5 & 3000.9 & 1848.5 & 2680.6  & 1232.1 & \textbf{466.44}          \\ \hline
		\textit{HPC2N}    & 297.18 & 426.99  & 609.77  & 157.71 & 118.01         & \textbf{117.01} \\ \hline
		\textit{Lublin-2} & 7842.5 & 9523.2 & 11265  & 787.89  & \textbf{698.34} & 724.51 \\ \hline
        \midrule
        
		\hline
		\multicolumn{7}{|c|}{\textit{Scheduling \textbf{with} Backfilling}}
		\\\hline
		\textit{Lublin-1} & 235.82 & 133.87 & 307.23 & 73.31  & 75.07  & \textbf{58.64}  \\ \hline
		\textit{SDSC-SP2} & 1595.1  & 1083.1 & 548.01 & 2167.8 & 1098.2    & \textbf{397.82} \\ \hline
		\textit{HPC2N}    & 127.38  & 97.39  & 175.12 & 122.04  & \textbf{71.95}  & 86.14 \\ \hline
		\textit{Lublin-2} & 247.61 & 318.35 & 379.59 & \textbf{91.99} & 148.25 & 118.79  \\ \hline
        \bottomrule
	\end{tabular}
	\end{center}
\end{table}

To summarize, we draw two conclusions from these results. First, \rbt{a heuristic} scheduler may perform well and poorly across different workloads. For example, with backfilling enabled, SJF performs the best on Lublin-2 ($91.99$) but performs the worst ($2167.8$) on SDSC-SP2. This shows the necessity of an adaptive scheduler, such as our RLScheduler, to work with different workloads. Second, across the listed workloads, RLScheduler is able to perform either comparably well to the best or is the best among the presented schedulers. These results conclude that, RLScheduler can adapt to different workloads with good performance.


\subsection{RLScheduler Performance on Different Goals}
This section reports RLScheduler performance on different optimization goals. We experimented on three metrics, including system resource utilization, average job slowdown and average job waiting time. We only report the results of resource utilization \rbt{in this section and leave the results of job slowdown and job waiting time in the Appendix}. Note that, in these evaluations, we used the same training settings as the previous ones. Only the scheduling metrics are different.

\subsubsection{Training on Resource Utilization}
\begin{figure}[h!]
	\centering
	\includegraphics[width=\linewidth]{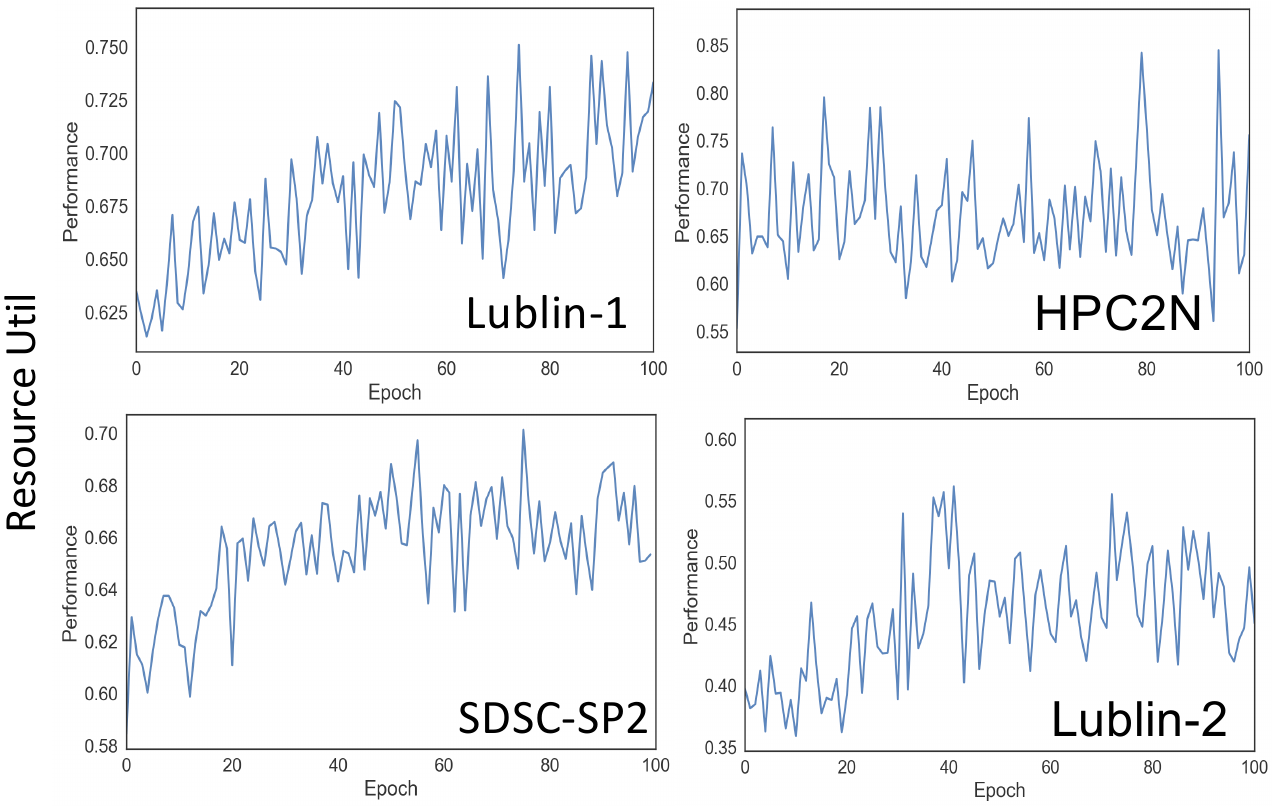}
	\caption{The training curves of RLScheduler in four different job traces targeting \textit{resource utilization}.}
	\label{exp:train-metrics}
\end{figure}

We first report the training curves of RLScheduler on this new metrics in Fig.~\ref{exp:train-metrics}. The results suggest that, RLScheduler still converges towards this new goal but with different patterns: there are more bumps during training. Moreover, certain workloads, such as HPC2N, seem to be improved slowly during the training (the peeks suggest the improvements). There are two potential reasons behind this. First, the HPC2N workload is much more uniformed regarding this metrics. Different schedulers actually do not change the utilization much (detailed results are shown in Table~\ref{tab:metrics}). So the learning of RL is slower with less knowledge from the training samples. Another reason is, in general system utilization has a narrow range, which makes the variance more noticeable for all training. 

\subsubsection{Scheduling on Resource Utilization}

\begin{table}[ht]
    \caption{Results of scheduling towards resource utilization.}
    \label{tab:metrics}
    \begin{center}
	\begin{tabular}{|c|c|c|c|c|c|c|}
		\hline
		\textit{Trace}     & FCFS   & WFP3    & UNICEP    & SJF    & F1     & \textbf{RL}     \\ \toprule
		\hline
		\multicolumn{7}{|c|}{\textit{Scheduling \textbf{without} Backfilling}}
		\\\hline
		\textit{Lublin-1} &0.657 & 0.747 & 0.691         & 0.762 & \textbf{0.816} & 0.714          \\ \hline
		\textit{SDSC-SP2} &0.670 & 0.658 & \textbf{0.688}         & 0.645 & 0.674          & 0.671 \\ \hline
		\textit{HPC2N}  &  0.638 & 0.636 & 0.636 & 0.640 & 0.637 &  \textbf{0.640}          \\ \hline
		\textit{Lublin-2} & 0.404 & 0.543 & 0.510 & 0.562 & 0.478           & \textbf{0.562} \\ \hline
        \midrule
        
		\hline
		\multicolumn{7}{|c|}{\textit{Scheduling \textbf{with} Backfilling}}
		\\\hline
		\textit{Lublin-1} & 0.868 & 0.864 & \textbf{0.883} & 0.778 & 0.840 & 0.850          \\ \hline
		\textit{SDSC-SP2} & 0.682          & 0.681 & 0.706 & 0.661 & 0.677 & \textbf{0.707} \\ \hline
		\textit{HPC2N}    & 0.639          & 0.637 & 0.638 & 0.641 & 0.638 & \textbf{0.642} \\ \hline
		\textit{Lublin-2} & 0.587  & 0.583 & 0.587 & 0.593 & 0.552 & \textbf{0.593} \\ \hline
        \bottomrule
	\end{tabular}
	\end{center}
\end{table}
Next, we explore the RLScheduler performance on scheduling different workloads targeting the new goal. We used  the same setting as the previous evaluations (repeated 10 times on randomly picked job sequences that contain 1024 continuous jobs). Table~\ref{tab:metrics} presents the results, in which the best scheduler for each job trace is \rbt{marked} bold. Note we have both results with or without backfilling reported here too.

\begin{table*}[ht]
    \caption{Performance comparisons of one RL-learned model (RL-$X$) being applied to other job traces (\textit{Y}).}
    \label{tab:cmp}
    \begin{center}
	\begin{tabular}{|c|c|c|c|c|c|c|}
    	\hline
    	\textit{Trace}     & Best Heuristic Sched   & Worst Heuristic Sched    & RL-\textit{Lublin-1}    & RL-\textit{SDSC-SP2}    & RL-\textit{HPC2N}     & RL-\textit{Lublin-2} \\ 
    	
    	\hline
		\multicolumn{7}{|c|}{\textit{Scheduling \textbf{without} Backfilling}}\\
		\hline
		\textit{Lublin-1} & 258.37 (F1) & 22274.74 (UNICEP) & \textbf{254.67}      & 482.62      & 283.00   & 334.73      \\ \hline
		
		\textit{SDSC-SP2} & 1232.06 (F1)  & 3000.88 (WFP3) & 1543.40     & \textbf{466.44}      & 1016.83  & 1329.41     \\ \hline
		
		\textit{HPC2N}    & \textbf{118.01  (F1)}          & 660.77 (UNICEP) & 169.91      & 300.43      & 186.42    & 236.00      \\ \hline
		
		\textit{Lublin-2} & 698.34 (F1)  & 11265.3 (UNICEP) & 665.49      & 805.16      & 648.52    & \textbf{724.51}      \\ \hline
		
		\textit{ANL Intrepid} & \textbf{8.39 (F1)}  & 35.11 (FCFS) & 9.91       & 9.61       & 8.93    & 9.75       \\ \hline
		\midrule
		
		\hline
		\multicolumn{7}{|c|}{\textit{Scheduling \textbf{with} Backfilling}}\\
		\hline
		\textit{Lublin-1} & 73.31 (SJF) & 307.23 (UNICEP) & 58.64 & 93.16 & \textbf{54.65} & 64.45        \\ \hline
		
		\textit{SDSC-SP2} & 548.01 (UNICEP)          & 2167.84 (SJF) & 1364.43 & \textbf{397.82} & 746.65 & 1192.97 \\ \hline
		
		\textit{HPC2N}    & \textbf{71.95 (F1)}          & 175.12 (UNICEP) & 115.93 & 128.73 & 115.79 & 144.54 \\ \hline
		
		\textit{Lublin-2} & \textbf{91.99 (SJF)}  & 379.59 (UNICEP) & 172.15 & 183.98 & 139.80 & 118.79 \\ \hline
		
		\textit{ANL Intrepid} & \textbf{2.73 (F1)}  & 4.12 (UNICEP) & 3.63 & 4.56 & 3.99 & 3.58 \\ \hline
		
        \bottomrule
	\end{tabular}
	\end{center}
\end{table*}

We have several observations from the results. First, \rbt{a heuristic scheduler that performs well on one goal may perform poorly on another goal even scheduling the same workload. For example, without backfilling, F1 performs the best for minimizing the \textit{average bounded slowdown} on Lublin-2 ($698.34$), but performs one of the worst in maximizing the \textit{system utilization} on the same job trace ($0.478$). This again motivates the need of RLScheduler, which can be adaptive towards different metrics.} Second, for this new optimization goal, RLScheduler still performs either comparably well to the best or is the best among the presented schedulers, showing its advantage over heuristic schedulers. Third, compared to \textit{bounded slowdown}, \textit{system utilization} is the more stable metrics. For some cases (e.g., HPC2N with backfilling), the performance differences among different schedulers are small. However, this does not suggest that system utilization is not important for job schedulers designs. The small change in the system utilization may lead to big difference in terms of the overall cost of the cluster. These results conclude that, RLScheduler can adapt to different optimization goals with good performance \rbt{(refer to Appendix section for results of other two scheduling metrics).}


\subsection{RLScheduler Stabilization}
One major concern about learning a batch job scheduler from trial-and-error on a job trace is whether the learned model would be too specific to the given job trace and can not handle even small shifts in the workloads.

This section addresses the RLScheduler stability concern. In particular, we experimented on applying the learned RL model (RL-\textit{X}) from job trace (\textit{X}) onto other job traces (\textit{Y}) and see how it would perform. Note that, these job traces have distinct characteristics as listed in Table~\ref{tab:jobtrace}. Hence, the evaluations result will show whether RLScheduler is sufficiently stable for production use. 


 Table~\ref{tab:cmp} presents the results, in which the best result for each case is marked bold. Similar to the previous performance evaluations, for each pair of RL-\textit{X} model and job trace \textit{Y}, we conducted the scheduling on 10 randomly sampled job sequences and reported their average. We only report the best and worst results of the heuristic schedulers (i.e., FCFS, SJF, WFP3, UNICEP, F1) due to the space limitation. Also, we only reported the result under the scheduling metrics \textit{average bounded slowdown}. Other metrics show similar results.

From these results, we observed that, a learned RL-\textit{X} model, regardless of which job trace it was trained based on, can be safely applied to other job traces \textit{Y}, without making catastrophic scheduling decisions. Although the performance will be degraded comparing with RL-\textit{X} on \textit{X}, its degradation is actually controlled: it will be no worse than using an inappropriate heuristic scheduler. Such a low-bound or stabilization makes RLScheduler practical in production systems.

\subsection{RLScheduler with Fairness}
\rbt{In previous evaluations, we show RLScheduler adapts well to each individual scheduling metric. But, in production system, it may require to consider multiple metrics at the same time, such as minimizing job slowdown and maximizing resource utilization together, or minimizing job slowdown and keeping fairness among users together. The fixed heuristic scheduling algorithms are clumsy to handle this. However, RLScheduler can still work via configuring its reward functions. We take fairness as an example to demonstrate it.}

\rbt{Fairness among users is a conjugated metrics, and can be applied upon other metrics to build per-user goals. Let's take `per-user average job slowdown' as an example. It means the scheduler needs to consider not only the average slowdown of all jobs, but also the average slowdown of each user's jobs. To optimize it, we should not starve one user nor slowdown all jobs for strictly enforcing the fairness.}

\rbt{To integrate fairness into RLScheduler, we change the reward function ($r$) from `average job slowdown of all jobs' to an aggregated value of `average job slowdowns of each user'.
The \textit{aggregator}
determines the way the scheduler would enforce the fairness. For example, we can use \textit{Maximal} as the aggregator, which means RLScheduler will focus on the user with maximal average job slowdown and learn to prioritize the user to minimize the overall maximal. In this way, it strives to enhance fairness and reduce the average job slowdown at the same time.} 

\rbt{\subsubsection{Evaluation Results} Due to the space limit, we only report the results of `bounded job slowdown with fairness' as an example to show the performance of RLScheduler. Because among our job traces, only SDSC-SP2 and HPC2N contain user information, we show the results of these two traces. We used \textit{Maximal} as the aggregator and conducted the evaluations in the same way as the previous ones (scheduled 10 randomly picked job sequences with 1024 jobs).}

\begin{table}[ht]
\centering
    \caption{Results of scheduling different job traces towards \textit{bounded job slowdown} with \textit{Maximal} \textit{Fairness}.}
    \label{tab:fairness}
    \begin{center}
	\begin{tabular}{|c|c|c|c|c|c|c|}
		\hline
		\textit{Trace}     & FCFS   & WFP3    & UNICEP    & SJF    & F1     & \textbf{RL}     \\ \toprule
		\hline
		\multicolumn{7}{|c|}{\textit{Scheduling \textbf{without} Backfilling}}
		\\\hline
		\textit{SDSC-SP2} &7257 & 14858 & 12234         & 12185 & 8260          & \textbf{4116} \\ \hline
		\textit{HPC2N}  &  2058 & 5107 & 5145 & 1255 & 1310 &  \textbf{1147}          \\ \hline
        \midrule
        
		\hline
		\multicolumn{7}{|c|}{\textit{Scheduling \textbf{with} Backfilling}}
		\\\hline
		\textit{SDSC-SP2} & 7356          & 8464 & 3840 & 10121 & 7799 & \textbf{2712} \\ \hline
		\textit{HPC2N}    & 1502          & 2125 & 2081 & 1491 & 583 & \textbf{519} \\ \hline
        \bottomrule
	\end{tabular}
	\end{center}
\end{table}

\rbt{The results in Table~\ref{tab:fairness} show that RLScheduler performs the best in both job traces after considering fairness. From these two traces, we observe that RLScheduler performs remarkably better than the best heuristic scheduler in SDSC-SP2 trace and only slightly better in HPC2N trace. The main reason of this difference is that, in HPC2N trace, jobs submitted from different users are highly unbalanced. For instance, one user ($u_{17}$) submitted around 40K jobs while the average number of jobs per-user is only 700. So, in a period of time, it is often that only one user or small number of users are submitting job, hence less impacted by the fairness.}

\rbt{In addition, RLScheduler can also work with quota-based fairness. In this case, RLScheduler's scheduling decisions that are violating user' quota will be masked illegal and ignored similarly to the case of insufficient resources.}

\subsection{RLScheduler Computational Cost}
\label{subsec:cost}
\rbt{We finally discuss the computational cost of RLScheduler.} There are mainly two computational parts in using RLScheduler: 1) train the model to learn the scheduling policy; 2) inference a learned model to generate a scheduling decision.

\begin{table}[ht]
	\caption{Computation cost of RLScheduler on CPU node}
	\label{tab:cost}
	\begin{center}
		\begin{tabular}{cr}
			\toprule
			Name & Time Cost\\
			\midrule
			SJF sorts 128 jobs and picks one& $0.71ms$\\
			RLScheduler DNN makes a decision & $0.30ms$\\
			\midrule
			RLScheduler DNN training (one epoch) & $123s$\\
			Converge on Lublin-1 & $1.1h$\\
			\bottomrule
		\end{tabular}
	\end{center}
\end{table}

We timed both computations on our evaluation platform (Intel Xeon Silver 4109T CPU and 32GB DDR4 DRAM) and presented the results in Table~\ref{tab:cost}. In summary, the trained RLScheduler DNN can make a decision for 128 pending jobs in $0.3ms$, compared to SJF sorting the same 128 jobs in $0.7ms$~\footnote{both implementations are based on python and can be improved}. The decision making of RLScheduler is comparably fast. In addition, such a time cost will not grow even when more jobs are pending in the system as more jobs will first be cut-off to \verb|MAX_OBSV_SIZE| (i.e., 128).


During RLScheduler training, one epoch takes around $123$ seconds and it typically takes less than 100 epochs to converge. Specifically, it took $1.1h$ to converge our training on Lublin-1 job trace. The computation will be much faster on GPU. But we consider this cost is acceptable as the training is only needed when the workload changes significantly or the metrics changes. Both of them typically will not happen hourly. 

	\section{Related Work}
	\label{sec:related}

	In HPC, batch job scheduling is a long-standing topic that draws lots of attentions in HPC community~\cite{tang2009fault,Carastan-Santos2017ObtainingLearning,Xhafa2010ComputationalProblems,Mu2001UtilizationBackfilling,agarwal2006heuristics,akyol2007review,al2012power,iosup2011performance}. In summary, researchers take various approaches and develop various scheduling policies, ranging from simple and classic policies (e.g., First Come First Served (FCFS) and Shortest Job First (SJF)) to complex policies (WFP3 and UNICEF)~\cite{tang2009fault}, from linear programming~\cite{al2012power,floudas2005mixed} to non-linear algorithms~\cite{hou1994genetic,singh1996mapping} and even neural networks~\cite{agarwal2006heuristics,akyol2007review}. Being different from this group of works, RLScheduler is built on deep reinforcement learning and is designed to be automated to the variations on both job loads and optimization goals. 
	
	Recently, reinforcement learning has also been studied and leveraged in various system optimization tasks. Examples include resource scheduling and task provisioning, such as DeepRM~\cite{Mao2016ResourceLearning,chen2017deep}, Decima~\cite{Mao2018LearningClusters}, and RRL~\cite{qin2019swift}; resource configuration tuning~\cite{tesauro2006hybrid,bu2012coordinated,padala2014scaling}; file system tuning~\cite{li2017capes} and performance prediction~\cite{hpdc17, pdsw19}. Although these studies leveraged reinforcement learning methods as RLScheduler does, they are not solving the automated batch job scheduling problem, hence miss the key training improvement and stabilization mechanisms proposed in RLScheduler.
	
	F1 and several priority functions from~\cite{Carastan-Santos2017ObtainingLearning} are considered as the state-of-the-art HPC batch job scheduler. These schedulers were built via non-linear regressing a large number of samples generated from brute force simulations. Compared with it, RLScheduler takes an unsupervised learning approach (reinforcement learning) and is accordingly automated and agile to different training loads and optimization goals and attains better performance in more cases.

	\section{Conclusion and Future Plan}
	\label{sec:conclude}
    This study presents RLScheduler, a deep reinforcement learning-based job scheduler. RLScheduler learns to schedule HPC batch jobs via its `trail and error' and is capable to learn high-quality scheduling policies for varying workloads and optimization goals. To prove the concept, we conducted extensive evaluations and confirmed that, RLScheduler performs well across different workloads and different optimization goals with high stability and reasonably low computation cost. Realizing RLScheduler is our first step. In the near future, we plan to \rbt{study more on multiple metrics optimization and} integrate it into real HPC cluster management tools such as Slurm. Moreover, we expect to apply the knowledge and experience we learn from this to other complex HPC settings. 
    
    \section*{Acknowledgment}
    We are thankful to the anonymous reviewers for their insightful feedback. This research is supported in part by the National Science Foundation under grants CNS-1852815, CNS-1817094, CNS-1817089, and MCB-1821828. 
    This research used resources of the Oak Ridge Leadership Computing Facility at the Oak Ridge National Laboratory, which is supported by the Office of Science of the U.S. Department of Energy under Contract No. DE-AC05-00OR22725.
    
    \rbt{
    \section*{Appendix}
    \label{sec:appendix}
    In this appendix section, we report the evaluation results of RLScheduler on other two metrics: \textit{average job slowdown} and \textit{average job waiting time}.

\subsection{RLScheduler on Average Job Slowdown}
The metrics of \textit{average slowdown} is very similar to the bounded slowdown, except that we calculate \textit{job slowdown} directly using the job's runtime without the "interactive thresholds". So short jobs with runtime close to $0$ may significantly increase its value. Fig.~\ref{exp:slowdown} reports the training curves of RLScheduler on different job traces. Comparing with the training curves of \textit{bounded slowdown} shown in Fig.~\ref{exp:train-workloads}, we observe similar convergence patterns, but with larger metrics values (affected by the short jobs).

\begin{figure}[h!]
	\centering
	\includegraphics[width=\linewidth]{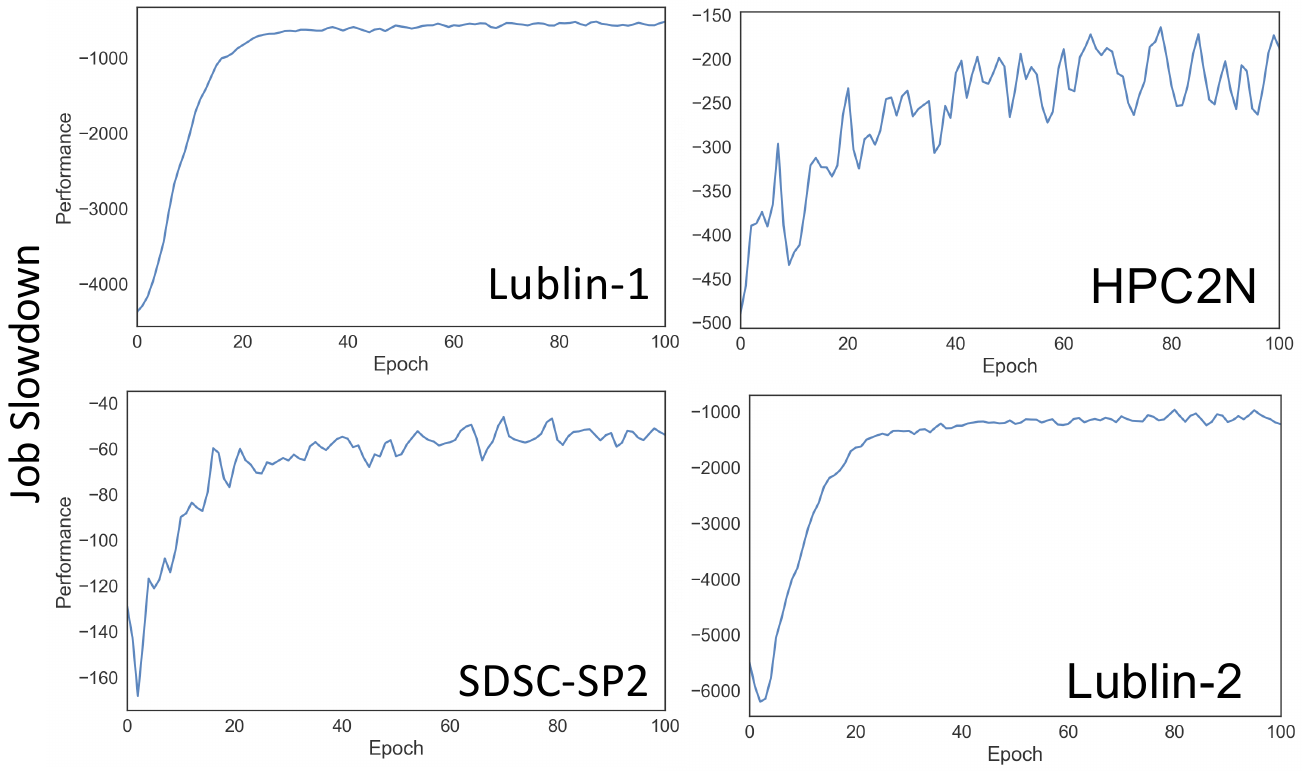}
	\caption{The training curves of RLScheduler in four different job traces targeting \textit{job slowdown}.}
	\label{exp:slowdown}
\end{figure}

In Table~\ref{tab:slowdown}, we further show the performance of RLScheduler when it actually schedules the workloads. We used the same setting as the previous evaluations (scheduled 10 randomly picked job sequences that contain 1024 jobs). From the results, we can observe that, although the existence of short jobs enlarges the metrics values, RLScheduler still performs either comparably well or better than other schedulers.

\begin{table}[ht]
\centering
    \caption{Results of scheduling towards \textit{Job Slowdown}.}
    \label{tab:slowdown}
    \begin{center}
	\begin{tabular}{|c|c|c|c|c|c|c|}
		\hline
		\textit{Trace}     & FCFS   & WFP3    & UNICEP    & SJF    & F1     & \textbf{RL}     \\ \toprule
		\hline
		\multicolumn{7}{|c|}{\textit{Scheduling \textbf{without} Backfilling}}
		\\\hline
		\textit{Lublin-1} &9711.1 & 26631 & 29995         & 397.3 & \textbf{339.3} & 429.5          \\ \hline
		\textit{SDSC-SP2} &2695.3 & 4548.6 & 1836.7         & 6790.8 & 2773.6          & \textbf{637.5} \\ \hline
		\textit{HPC2N}  &  883.0 & 1349.5 & 2137.3 & 326.4 & 273.3 &  \textbf{270.7}          \\ \hline
		\textit{Lublin-2} & 13914 & 18948 & 21978         & 1489.7 & 1925.4           & \textbf{1181.2} \\ \hline
        \midrule
        
		\hline
		\multicolumn{7}{|c|}{\textit{Scheduling \textbf{with} Backfilling}}
		\\\hline
		\textit{Lublin-1} & 316.72 & 199.67 & 478.21 & 91.82 & 87.37 & \textbf{85.55}          \\ \hline
		\textit{SDSC-SP2} & 3952.4          & 2519.2 & 870.4 & 5289.8 & 2670.5 & \textbf{619.5} \\ \hline
		\textit{HPC2N}    & 312.3          & 355.1 & 755.6 & 217.1 & \textbf{123.5} & 170.7 \\ \hline
		\textit{Lublin-2} & 275.6  & 310.1 & 423.6 & 162.6 & 133.2 & \textbf{114.3} \\ \hline
        \bottomrule
	\end{tabular}
	\end{center}
\end{table}

\subsection{RLScheduler on Average Job Waiting Time}
We repeated similar experiments with the scheduling metrics as minimizing the \textit{average job waiting time} and report the results here. Fig.~\ref{exp:waittime} shows the training curves of RLScheduler on different job traces. As the job waiting time may be huge, the metrics values in the vertical axis also become much larger. But, we can still observe similar, fast convergence patterns as seen in the previous evaluations, showing the generality of RLScheduler.

\begin{figure}[h!]
	\centering
	\includegraphics[width=\linewidth]{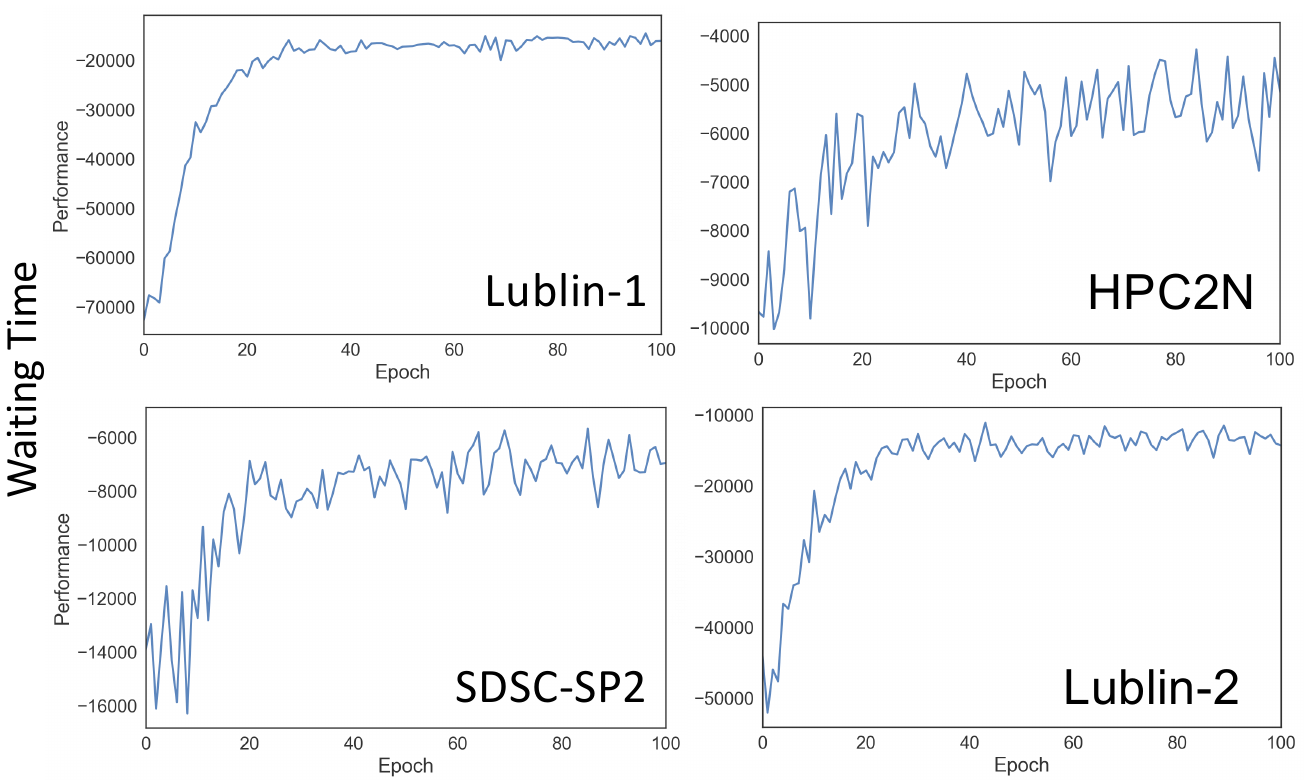}
	\caption{The training curves of RLScheduler in four different job traces targeting \textit{job Waiting Time}.}
	\label{exp:waittime}
\end{figure}

Table~\ref{tab:waittime} shows the scheduling results. Again, the setting is the same as previous evaluations (scheduled 10 randomly picked job sequences that contain 1024 jobs). From these results, we observe that RLScheduler still performs either the best or close to the best schedulers across all workloads.

\begin{table}[h!]
\centering
    \caption{Results of scheduling towards \textit{Job Waiting Time}.}
    \label{tab:waittime}
    \begin{center}
	\begin{tabular}{|c|c|c|c|c|c|c|}
		\hline
		\textit{Trace}     & FCFS   & WFP3    & UNICEP    & SJF    & F1     & \textbf{RL}     \\ \toprule
		\hline
		\multicolumn{7}{|c|}{\textit{Scheduling \textbf{without} Backfilling}}
		\\\hline
		\textit{Lublin-1} &241111 & 476218 & 683473         & 42509 & \textbf{18936} & 20550          \\ \hline
		\textit{SDSC-SP2} &57850 & 85358 & 103734         & 46382 & 26146          & \textbf{18984} \\ \hline
		\textit{HPC2N}  &  13938 & 23928 & 29633 & 7071 & 6865 &  \textbf{6858}          \\ \hline
		\textit{Lublin-2} & 134542 & 178924 & 202814         & 18278 & 26104           & \textbf{17809} \\ \hline
        \midrule
        
		\hline
		\multicolumn{7}{|c|}{\textit{Scheduling \textbf{with} Backfilling}}
		\\\hline
		\textit{Lublin-1} & 24887 & 19112 & 32110 & 30099 & 12692 & \textbf{12460}          \\ \hline
		\textit{SDSC-SP2} & 32577          & 26325 & 25392 & 34834 & 21830 & \textbf{13817} \\ \hline
		\textit{HPC2N}    & 7184          & 9154 & 11855 & 5760 & \textbf{4633} & 5394 \\ \hline
		\textit{Lublin-2} & 6300  & 8713 & 8155 & \textbf{4011} & 5828 & 5172 \\ \hline
        \bottomrule
	\end{tabular}
	\end{center}
\end{table}

	}

	\bibliographystyle{IEEEtran}
	{
		\bibliography{bib}
	}
\end{document}